\documentclass[11pt,a4paper]{article}
\usepackage{jcappub}

\usepackage{epsfig,psfrag,graphics,verbatim}
\usepackage{graphicx}

\usepackage{dcolumn}
\usepackage{bm}
\usepackage{color}
\usepackage{ulem}
\newcommand{\params}{{\mathbf \Theta}}

\newcommand{\loc}{\text{loc}}
\newcommand{\esc}{\text{esc}}
\newcommand{\lsr}{\text{lsr}}
\newcommand{\tmin}{\text{min}}
\newcommand{\Nobs}{\hat{N}}

\newcommand{\be}{\begin{equation}}
\newcommand{\ee}{\end{equation}}
\newcommand{\sigmaSI}{\sigma_{\chi-p}^\text{SI}}
\newcommand{\sigmaSD}{\sigma_{\chi-p}^\text{SD}}

\newcommand{\BR}{BR}
\newcommand\RBtaunu{\frac{\BR(B_u \to \tau \nu)}{\BR(B_u \to \tau \nu)_{SM}}}
\newcommand\DeltaO{\Delta_{0-}}
\newcommand\RBDtaunuBDenu{\frac{\BR(B \to D \tau \nu)}{\BR(B \to D e \nu)}}
\newcommand\Rl{R_{l23}}

\newcommand\Dstaunu{\BR(D_s \to \tau \nu)}
\newcommand\Dsmunu{\BR(D_s\to \mu \nu)} 
\newcommand\Dmunu{\BR(D \to \mu \nu)} 
\newcommand\brbsmumu{\BR(\overline{B}_s\to\mu^+\mu^-)}
\newcommand{\amusm}{a_{\mu}^{\text{SM}}}
\newcommand{\gmtwo}{(g-2)_{\mu}}
\newcommand{\hl}{h}
\newcommand\mhl{m_h}
\newcommand{\zetah}{\zeta_h}
\newcommand{\brbsgamma}{BR(\bar{B} \rightarrow X_s\gamma) }

\newcommand{\gev}{\mbox{ GeV}}

\newcommand{\cl}{\text{CL}}

\begin{document}


\title{Global fits of the cMSSM including the first LHC and XENON100 data}

\author[a]{Gianfranco Bertone}
\author[b]{David G. Cerde\~{n}o}
\author[c,1]{Mattia Fornasa \note{MultiDark fellow}}
\author[d]{Roberto Ruiz de Austri}
\author[e]{Charlotte Strege}
\author[e,f]{Roberto Trotta}

\affiliation[a]{GRAPPA Institute, University of Amsterdam, Science Park 904, 1090 GL Amsterdam, Netherlands}
\affiliation[b]{Departamento de F\'{\i}sica Te\'orica and Instituto de F\'{\i}sica Te\'orica UAM-CSIC, Universidad Aut\'onoma de Madrid, Cantoblanco, E-28049 Madrid, Spain}
\affiliation[c]{Instituto de Astrof\'{\i}sica de Andaluc\'{\i}a (CSIC), E-18008, Granada, Spain}
\affiliation[d]{Instituto de F\'isica Corpuscular, IFIC-UV/CSIC, Valencia, Spain} 
\affiliation[e]{Astrophysics Group, Imperial College London, Blackett Laboratory, Prince Consort Road, London SW7 2AZ, UK}
\affiliation[f]{African Institute for Mathematical Sciences, 6 Melrose Rd, Muizenberg, 7945, Cape Town, South Africa}

\abstract{
We present updated global fits of the constrained Minimal Supersymmetric Standard Model (cMSSM), including the most recent constraints from the ATLAS and CMS detectors at the LHC, as well as the most recent results of the XENON100 experiment. Our robust analysis takes into account both astrophysical and hadronic uncertainties that enter in the calculation of the rate of WIMP-induced recoils in direct detection experiment.
We study the consequences for neutralino Dark Matter, and show that current direct detection data already allow to robustly rule out the so-called Focus Point region, therefore demonstrating the importance of particle astrophysics experiments in constraining extensions of the Standard Model of Particle Physics. We also observe an increased compatibility between results obtained from a Bayesian and a Frequentist statistical perspective. 
We find that upcoming ton-scale direct detection experiments will probe essentially the entire currently favoured region (at the 99\% level), almost independently of the statistical approach used. Prospects for indirect detection of the cMSSM are further reduced.}



\arxivnumber{1107.1715}

\maketitle

\section{Introduction}
\label{secintro}
The constrained Minimal Supersymmetric Standard Model (cMSSM) \cite{Chamseddine:1982jx,Kane:1993td} is the most widely discussed extension of the Standard Model (SM) of particle physics. Despite its relative simplicity, this model has the advantage of capturing some key phenomenological feature of Supersymmetry (SUSY), while making definite predictions for the properties of the the lightest neutralino $\tilde{\chi}_1^0$ (which we denote $\chi$ here, for simplicity), a linear superposition of the superpartners of the neutral  Gauge bosons and the neutral Higgses, which is by far the most popular Dark Matter (DM) candidate \cite{Bertone:2010zz,Bergstrom:2000pn,Munoz:2003gx,Bertone:2004pz}.

Recently, the ATLAS and CMS experimental collaborations have published the first search for supersymmetry (SUSY) \cite{Chatrchyan:2011wc,Khachatryan:2011tk,daCosta:2011qk,Aad:2011ks,Aad:2011xm}, based on  $\sqrt{s} = 7$ TeV proton-proton collisions at the LHC recorded during 2010 (see the experimental papers above for details on the event topologies, selection cuts, etc.). Since no excess above the SM predictions has been observed, these searches have set new interesting constraints on physics beyond the SM, and several groups of authors have already studied the impact of these results on the cMSSM and more in general on SUSY (see e.g. Refs. \cite{Buchmueller:2011aa,Allanach:2011ut,Akula:2011dd,Strumia:2011dv,Allanach:2011wi,Buchmueller:2011ki,Farina:2011bh,Strumia:2011dy}).  

Here, we perform new global fits of the cMSSM, following the methodology outlined in Refs. \cite{deAustri:2006pe,Trotta:2008bp,Feroz:2011bj}, where LEP constraints, precision tests of the SM and the most recent cosmological constraint on the relic abundance of the DM were implemented using nested sampling as a scanning algorithm, in order to derive both the most probable regions of the cMSSM (Bayesian) and the best fit regions (frequentist). We update the analysis to include new $B$-physics observables, as well as both the aforementioned LHC constraints and the upper limit on the WIMP-nucleon scattering cross-section arising from DM direct detection experiments (see e.g. Ref.~\cite{Cerdeno:2010jj} and references therein). We implement the findings of the XENON100 collaboration \cite{xenon:2011hi}, which recently reported the results of a two-phase time projection chamber with 48 kg of liquid xenon as ultra-low background fiducial target. In 100.9 live days of data, acquired between January and June 2010, the collaboration recorded three candidate events, with an expected background of 1.8 $\pm$ 0.6 events.
The rate of DM-induced recoil events in a direct detection
experiment being directly proportional to the spin-independent scattering cross section of neutralinos 
off protons $\sigmaSI$, the upper limit on the recoil rate translates into an upper limit on $\sigmaSI$. 

The predictions for $\sigmaSI$ within the cMSSM (given constraints other than direct detection data) span
approximately the range between $10^{-7}$ pb to $10^{-10}$ pb, and since XENON100 excludes  $\sigmaSI$ larger than $7.0 \times 10^{-9}$ pb for a DM particle mass of 50 GeV at the 90\% confidence level, one sees immediately that this has the potential to rule out significant portions of the cMSSM, under the hypothesis of a standard expansion rate of the Universe, and that the neutralinos make up most of the DM in the Universe.

Particular care must be paid, however, when translating the information 
provided by a direct detection experiment to constraints on the cMSSM 
parameter space, since this requires specific modeling of the velocity 
distribution of DM particles and of the DM density in the solar neighborhood. 
As a consequence, the upper limit from XENON100 can be considered robust
only if the uncertainties on these quantities are taken into account.
Moreover, the value of $\sigmaSI$ depends on nuclear
physics quantities, namely the hadronic matrix elements, which parametrize the quark composition of the proton, and this affects the way direct detection data are applied to constrain cMSSM parameters.
Therefore in the present analysis we incorporate the local density, the quantities
defining the DM velocity distribution and those modeling the proton 
composition as nuisance parameters, which are then either marginalized 
(i.e., integrated out) or maximised over to obtain inferences on the cMSSM 
parameters that fully account for those astrophysical and hadronic 
uncertainties.

The paper is organized as follows: in Section~\ref{sec:theory} we present our 
theoretical framework, in Section~\ref{secDD} we present the formalism for 
the computation of recoil events, our modelization of the DM velocity 
distribution and our implementation of the XENON100 data. Section \ref{secresults} is 
devoted to the presentation of the results, while prospects for the discovery of SUSY in the cMSSM are discussed in Section \ref{sec:prospects}. We conclude in 
Section \ref{secconclusion}.

\section{Theoretical and statistical framework}
\label{sec:theory}

\subsection{Model, priors and scanning algorithm}

The five parameters defining a generic cMSSM are $i)$ the coefficient of 
the universal mass term for scalars $m_0$ and $ii)$ for gauginos $m_{1/2}$, 
$iii)$ the coefficient of the trilinear interaction $A_0$, $iv)$ the ratio 
between the vacuum expectation values of the Higgs bosons $\tan\beta$ and 
$v)$ the sign of the mass term for Higges sgn$(\mu)$.
We consider only four of these in our scan, fixing the sign
of the $\mu$ coefficient to be positive, a choice favoured by the results
on the anomalous magnetic moment of the muon.
The scanned range of the remaining four cMSSM parameters is summarized in
Table~\ref{tab:cMSSM_params}. Moreover, since residual experimental 
uncertainties on the value of some SM quantities can have strong effects on 
the values for the observables we are interested in~\cite{Roszkowski:2007fd}, 
we also include in our scan the four SM parameters indicated in 
Table~\ref{tab:nuis_params}, which are considered as nuisance parameters.

\begin{table}
\begin{center}
\begin{tabular}{l l l}
\hline
\hline
\multicolumn{3}{c}{cMSSM Parameters} \\
 & Flat priors & Log priors \\
\hline
$m_0$ [GeV] & (50.0, 4000.0) & ($10^{1.7}$, $10^{3.6}$) \\
$m_{1/2} $ [GeV] & (50.0, 4000.0) & ($10^{1.7}$, $10^{3.6}$) \\
$A_0$ [GeV] & \multicolumn{2}{c}{(-4000.0, 4000.0)} \\
$\tan\beta$ & \multicolumn{2}{c}{(2.0, 65.0)} \\
\hline
\end{tabular}
\end{center}
\caption{\fontsize{9}{9}\selectfont cMSSM parameters and the range covered on the scan. Flat priors are uniform in the masses, while log priors are uniform in the logarithm of the masses.}\label{tab:cMSSM_params}
\end{table}

\begin{table}
\begin{center}
\begin{tabular}{l l l l }
\hline
\hline
\multicolumn{4}{c}{SM nuisance parameters} \\
\hline
 & Gaussian prior  & Range scanned & Ref. \\
\hline
$M_t$ [GeV] & $173.1 \pm 1.3$  & (167.0, 178.2) &  \cite{topmass:09} \\
$m_b(m_b)^{\bar{MS}}$ [GeV] & $4.20\pm 0.07$ & (3.92, 4.48) &  \cite{pdg07}\\
$[\alpha_{em}(M_Z)^{\bar{MS}}]^{-1}$ & $127.955 \pm 0.030$ & (127.835, 128.075) &  \cite{pdg07}\\
$\alpha_s(M_Z)^{\bar{MS}}$ & $0.1176 \pm  0.0020$ &  (0.1096, 0.1256) &  \cite{Hagiwara:2006jt}\\
\hline
\multicolumn{4}{c}{Astrophysical nuisance parameters} \\
\hline
$\rho_{\loc}$ [GeV/cm$^3$] & $0.4\pm 0.1$ & (0.001, 0.900) & \cite{Pato:2010zk}\\
$v_{\lsr}$ [km/s] & $230.0 \pm 30.0$ &  (80.0, 380.0) & \cite{Pato:2010zk}\\
$v_{\esc}$ [km/s] & $544.0 \pm 33.0$ & (379.0, 709.0) & \cite{Pato:2010zk}\\
$v_d$ [km/s] & $282.0 \pm 37.0$ & (98.0, 465.0) & \cite{Pato:2010zk}\\
\hline
\multicolumn{4}{c}{Hadronic nuisance parameters} \\
\hline
$f_{Tu}$ & $0.02698 \pm 0.002$ & (0.010, 0.045) &  \cite{Ellis}\\
$f_{Td}$ & $0.03906 \pm 0.00395$ &  (0.015, 0.060) &  \cite{Ellis}\\
$f_{Ts}$ & $0.363 \pm 0.119$ & (0.000, 0.85) &  \cite{Ellis}\\
\hline
\hline
\end{tabular}
\end{center}
\caption{\fontsize{9}{9}\selectfont Nuisance parameters adopted in the scan of the cMSSM parameter space, indicating the mean and standard deviation adopted for the Gaussian prior on each of them, as well as the range covered in the scan. For scans in which the hadronic and/or astrophysical nuisance parameters have been held fixed, their value corresponds to the mean indicated here.}\label{tab:nuis_params} 
\end{table}

The cMSSM parameters together with the nuisance parameters in our scan are indicated by
$\params$.
The core of Bayesian statistics is Bayes' theorem, which reads (see Ref.~\cite{Trotta:2008qt} for further details):
\begin{equation}
p(\params|\mathbf{D})=
\frac{p(\mathbf{D}|\params) p(\params)}{p(\mathbf{D})},
\label{eqn:Bayes}
\end{equation}
where $\mathbf{D}$ is the set of experimental data used to 
constraint the parameter space, while $p(\params|\mathbf{D})$ on the 
l.h.s. of Eq.~\eqref{eqn:Bayes} is the posterior probability density function (pdf),
$p(\mathbf{D}|\params)$ is the likelihood function (also
indicated as $\mathcal{L}(\params)$) and 
$p(\params)$ is the prior pdf, representing our knowledge of the
parameters before the data is taken into account. Bayes' theorem 
expresses how this prior information is updated by means of the experimental 
data through the likelihood function, so that the final pdf for 
$\params$ depends both on the priors and on the likelihood. The evidence (or model likelihood) $p(\mathbf{D})$ only depends on the data so, for our purposes it represents a normalization
factor and can thus be dropped. Different choices can be made for the prior 
probability in Eq.~\eqref{eqn:Bayes}. However if the pdf at the end still
exhibits a residual dependence on the priors, this represents a clear sign 
that the experimental data employed are not constraining enough to overcome
the priors and, as a consequence, the resulting pdf should be considered
with care~\cite{Trotta:2008bp,Scott:2009jn}.
Following Refs.~\cite{Trotta:2008bp} and \cite{Feroz:2011bj}, in order to assess the robustness of our inference we consider two sets of priors: one that is uniform in the cMSSM masses (``flat'' priors) and one that is uniform in their log (``log'' priors). Both sets of priors are uniform on $A_0$ and $\tan\beta$, and the ranges are indicated in Table~\ref{tab:cMSSM_params}. For the SM, astrophysical and hadronic nuisance parameters our priors represents current available measurements of those quantities, and are indicated in Table~\ref{tab:nuis_params} (further details are given in Section~\ref{sec:nuisance} below). Furthermore, we also present results for the profile likelihood, which is in principle independent of the prior~\cite{Feroz:2011bj}. 

When presenting results for 1-dimensional or 2-dimensional subsets of the parameter space, the
remaining parameters can be eliminated in two different ways:
\begin{itemize}
\item By marginalizing over them (i.e., integrating over -- Bayesian), resulting in the 1-dimensional (or 2-dimensional)
posterior pdf for the $i$-th parameter:
\begin{equation}
p(\Theta_i|\mathbf{D})=\int p(\params|\mathbf{D}) 
d\Theta_1 ... d\Theta_{i-1} d\Theta_{i+1} d\Theta_{n},
\end{equation}
\item By maximizing over them (i.e., profiling over the likelihood -- Frequentist), resulting in the 1-dimensional (or 2-dimensional)
profile likelihood function:
\begin{equation}
{\mathcal L}(\Theta_i)= \max_{\Theta_1,...,\Theta_{i-1},\Theta_{i+1},...,\Theta_{n}}
\mathcal{L}(\params).
\end{equation}
\end{itemize}
Inferences resulting from those two approaches have, in general, a different meaning and may produce
different results: the definition of the profile likelihood is more suited 
to determine possible small regions in the parameter space with large 
likelihood values, while, on the contrary, the pdf, integrating over 
hidden directions, correctly accounts for volume effects. Usually the largest 
amount of information on the structure of the scanned space derives from
the comparison of the two~\cite{Scott:2009jn,Feroz:2011bj}.
For both choices of priors, we use a modified version of the public \texttt{SuperBayeS v1.5.1} package \cite{SuperBayeS}\footnote{For this paper, the public \texttt{SuperBayeS v1.5.1} code has been modified to interface with with 
DarkSUSY 5.0.5~\cite{DarkSUSY,Gondolo:2004sc}, SoftSUSY 2.0.18 
\cite{SoftSUSY,Allanach:2001kg}, MicrOMEGAs 2.0  
\cite{MicrOMEGAs,Belanger:2006is}, SuperIso 2.4 \cite{SuperIso,Mahmoudi:2008tp} and SusyBSG 1.4 \cite{SusyBSG,Degrassi:2007kj}.} to obtain posterior samples of Eq.~\eqref{eqn:Bayes}, adopting
MultiNest v2.8~\cite{Feroz:2007kg,Feroz:2008xx} as a scanning algorithm. We use running parameters as recommended in Ref.~\cite{Feroz:2011bj}, namely a number of live points $n_\text{live} = 20,000$ and a tolerance parameter $\text{tol} = 10^{-4}$, which are tuned to obtain an accurate evaluation of the profile likelihood. Our final inferences for each of the log and flat priors are obtained from chains generated with approximately $13$ ($41$) million likelihood evaluations for scans with fixed (varying) astrophysical and hadronic nuisance parameters, with an overall efficiency of about 5\%. The profile likelihood plots are derived from merged chains from both of our log and flat priors scans, as advocated in Ref.~\cite{Feroz:2011bj}, and are obtained from over 1 million samples. 

\subsection{The likelihood function} 

The experimental data $\mathbf{D}$ included in the likelihood are summarized in Table~\ref{tab:SM_constraints}. Observable quantities already included in Ref.~\cite{Trotta:2008bp} have been used here with the same likelihood function as in Ref.~\cite{Trotta:2008bp} (for the {\ttfamily ALL} setup), with updates to central values and standard deviations where applicable as indicated in Table~\ref{tab:SM_constraints}. In particular, we include LEP bounds on the Higgs mass, precision tests of the SM and an updated constraint on the DM relic density from WMAP 7-years data (assuming that neutralinos are the sole constituent of DM). We have also made the following modifications with respect to the treatment in Ref.~\cite{Trotta:2008bp}: the $\brbsgamma$ is now computed with the 
numerical code \texttt{SusyBSG}~\cite{Degrassi:2007kj}, which uses the full 
NLO QCD contributions, including the two-loop calculation of the gluino 
contributions presented in Ref.~\cite{Degrassi:2006eh} and the results of Ref.~\cite{D'Ambrosio:2002ex} for the remaining non-QCD $\tan\beta$-enhanced 
contributions\footnote{For certain choices of the input parameters 
it may happen that the masses of two particles are accidentally close 
to each other. In some of these cases, this leads to numerical instabilities 
which usually are circumvented by compiling the program to quadruple 
precision. However, we have observed that in rare occasions numerical instabilities may persist.}.
The other $B(D)$-physics observables summarized in Table~\ref{tab:SM_constraints} have been computed with the code \texttt{SuperIso} (for details
on the computation of the observables see Ref.~\cite{Mahmoudi:2008tp} and 
references therein). The anomalous magnetic moment of the muon, $\delta a_{\mu}^{\text{SUSY}}$, is also computed with~\texttt{SuperIso}. Both \texttt{SuperIso} and  \texttt{SusyBSG} have been integrated in a modified version of \texttt{SuperBayeS v1.5.1}.

In addition, we also include the recent constraints from SUSY searches at the LHC. We follow the analysis presented by the ATLAS collaboration in 
Ref.~\cite{Aad:2011hh}, since it is the one associated to the more stringent 
constraints: a data sample corresponding to a luminosity of 35 pb$^{-1}$ for a 
center-of-mass energy of $\sqrt{s}=7$ GeV is used to search for a SUSY signal
in events with an isolated electron or muon with high transverse momentum,
at least three hadronic jets and missing transverse momentum.
No significant deviations are found with respect to SM predictions, and this
is used to derive the constraints on the ($m_0,m_{1/2}$) plane indicated
in Fig.~2 of Ref.~\cite{Aad:2011hh} as a solid red line. We note that while this 95\% 
exclusion line has been obtained assuming values of $\tan\beta = 3.0$ and $A_0 = 0$ GeV, the final result is not expected 
to change much for different values of these two parameters since it is 
obtained from decay channels that are fairly independent on the value of 
$\tan\beta$ and $A$. We include those exclusion limits by assigning a likelihood equal to zero to samples falling below the exclusion line. 

The likelihood function for the direct detection constraints is a central ingredient of this work, and is presented in more detail in the next section.

\begin{table*}
\begin{center}
\begin{tabular}{|l | l l l | l|}
\hline
\hline
Observable &   Mean value & \multicolumn{2}{c|}{Uncertainties} & Ref. \\
 &   $\mu$      & ${\sigma}$ (exper.)  & $\tau$ (theor.) & \\\hline
$M_W$ [GeV] & 80.398 & 0.025 & 0.015 & \cite{lepwwg} \\
$\sin^2\theta_{eff}$ & 0.23153 & 0.00016 & 0.00015 & \cite{lepwwg} \\
$\delta a_\mu^\text{SUSY} \times 10^{10}$ & 29.6 & 8.1 & 2.0 & \cite{Davier:tau10} \\
$\brbsgamma \times 10^4$ & 3.55 & 0.26 & 0.30 & \cite{hfag}\\
$\Delta M_{B_s}$ [ps$^{-1}$] & 17.77 & 0.12 & 2.40 & \cite{cdf-deltambs} \\
$\RBtaunu$   &  1.28  & 0.38  & - & \cite{hfag}  \\
$\DeltaO  \times 10^{2}$   &  3.6  & 2.65  & - & \cite{:2008cy}  \\
$\RBDtaunuBDenu \times 10^{2}$ & 41.6 & 12.8 & 3.5  & \cite{Aubert:2007dsa}  \\
$\Rl$ & 1.004 & 0.007 & -  &  \cite{Antonelli:2008jg}  \\
$\Dstaunu \times 10^{2}$ & 5.38 & 0.32 & 0.2  & \cite{hfag}  \\
$\Dsmunu  \times 10^{3}$ & 5.81 & 0.43 & 0.2  & \cite{hfag}  \\
$\Dmunu \times 10^{4}$  & 3.82  & 0.33 & 0.2  & \cite{hfag} \\
$\Omega_\chi h^2$ & 0.1123 & 0.0035 & 10\% & \cite{Jarosik:2010iu} \\\hline\hline
   &  Limit (95\%~\cl)  & \multicolumn{2}{r|}{$\tau$ (theor.)} & Ref. \\ \hline
$\brbsmumu$ &  $ <5.8\times 10^{-8}$
 & & 14\% & \cite{cdf-bsmumu}\\
$\mhl$  & $>114.4\gev$\ (SM-like Higgs)  
 & & 3 $\gev$ & \cite{lhwg} \\
$\zetah^2$
& $f(m_h)$\ (see Ref.~\cite{deAustri:2006pe})  & \multicolumn{2}{r|}{negligible}  & \cite{lhwg} \\
$m_{\tilde{q}}$ & $>375$ GeV  & & 5\% & \cite{pdg07}\\
$m_{\tilde{g}}$ & $>289$ GeV  & & 5\% & \cite{pdg07}\\
other sparticle masses  &  \multicolumn{3}{c|}{As in table~4 of
  Ref.~\cite{deAustri:2006pe}.}  & \\
$m_0, m_{1/2}$ & \multicolumn{3}{c|}{LHC exclusion limits, see text} & \cite{Aad:2011hh} \\
$m_\chi - \sigmaSI$ & \multicolumn{3}{c|}{XENON100 exclusion limits, see text} &\cite{xenon:2011hi} \\
\hline
\end{tabular}
\end{center}
\caption{\fontsize{9}{9} \selectfont Summary of the observables used
for the computation of the likelihood function
Upper part: Observables for which a positive measurement has been made. 
$\delta a_\mu^\text{SUSY}= a_\mu^{\rm exp}-\amusm$ denotes the discrepancy between
the experimental value and the SM prediction of the anomalous magnetic
moment of the muon $\gmtwo$.
For each quantity we use a
likelihood function with mean $\mu$ and standard deviation $s =
\sqrt{\sigma^2+ \tau^2}$, where $\sigma$ is the experimental
uncertainty and $\tau$ represents our estimate of the theoretical
uncertainty. Lower part: Observables for which only limits currently
exist. The explicit form of the likelihood function is given in
Ref.~\cite{deAustri:2006pe}, including in particular a smearing out of
experimental errors and limits to include an appropriate theoretical
uncertainty in the observables. $\mhl$ stands for the light Higgs mass
while $\zetah^2\equiv g^2(\hl ZZ)_{\text{MSSM}}/g^2(\hl ZZ)_{\text{SM}}$,
where $g$ stands for the Higgs coupling to the $Z$ and $W$ gauge boson
pairs. \label{tab:SM_constraints}}
\end{table*}

\section{Constraints from direct detection data}
\label{secDD}

\subsection{Calculation of the recoil rate} 

Before discussing our implementation of direct detection results in the likelihood, we briefly review the calculation of the recoil rate, with particular emphasis on the astrophysical uncertainties (see also the discussion in 
Refs.~\cite{Pato:2010zk,Bertone:2010rv,ST}). 

The rate of recoil events produced by DM interactions 
with a nucleus of mass $m_N$ is given by 
\begin{equation}
\frac{{\rm d}R}{ {\rm d}E} =
\frac{\rho_{\loc}}{m_\chi m_N}
 \int_{\mathcal{V}} d\mathbf{v} \, vf(\mathbf{v+v}_\text{Earth}) 
\frac{{\rm d}\sigma_{\chi-N}}{ {\rm d}E}(v,E).
\label{eqn:diff_rate}
\end{equation}
Here, $\rho_{\loc}$ is the local DM density, $\mathbf{v}$ the DM velocity in
the detector rest frame, $\mathbf{v}_\text{Earth}$ the Earth velocity in the
rest frame of the Galaxy, $f(\mathbf{w})$ the DM velocity distribution
function and ${\rm d} \sigma_{\chi-N}/{\rm d}E$ is the differential cross section 
for the interaction between the neutralino and the nucleus.
The integral is over the volume $\mathcal{V}$ of phase space for which $v$ is smaller than the escape velocity $v_{\esc}$ and larger than the minimal 
velocity $v_{\tmin}(E)$ able to produce a recoil with energy $E$: 
$v_{\tmin}(E)=\sqrt{m_N E/2 \mu_N^2}$, where 
$\mu_N=m_\chi m_N (m_\chi+m_N)^{-1}$ is the neutralino-nucleus reduced mass.

We will focus here only on the case of spin independent (SI) interaction, 
assuming also that protons and neutrons are characterized by the same cross 
section. Therefore:
\begin{equation}
\frac{{\rm d}\sigma_{\chi-N}}{{\rm d} E}(v,E)=\frac{m_N}{2\mu_p^2 v^2} 
\sigmaSI \sum_i \omega_i A_i^2 F_i^2(A_i,E),
\label{eqn:diff_cross_section}
\end{equation}
where $\mu_p$ is the neutralino-proton reduced mass. The sum is over all 
the isotopes present in the detector, each one with its own abundance factor
$\omega_i$. $A_i$ is the atomic mass number of the $i$-th nuclear species
and $F(A_i,E)$ the nuclear form factor, which we have taken from Ref.~\cite{Duda:2006uk}.

\subsection{Astrophysical and hadronic nuisance parameters}
\label{sec:nuisance}

Regarding the velocity distribution function entering Eq.~\eqref{eqn:diff_rate}, we account for uncertainties in the halo model by considering the following parametrization:
\begin{equation}
f(\mathbf{v})=f_0  
\left[ \exp \left( \frac{3 (v_{\esc}^2-v^2)}{2v_d^2} \right) -1 \right]^k.
\label{eqn:vel_distribution}
\end{equation}
The distribution function depends on three parameters, $(k,v_{\esc},v_d)$, and
$f_0$ is simply a normalization factor.
Eq.~\eqref{eqn:vel_distribution} has been introduced in 
Ref.~\cite{Lisanti:2010qx} as an {\it Ansatz} able to reproduce the phase 
space structure of $N$-body simulated DM halos and, compared to the standard
Maxwellian distribution, it predicts a smaller number of DM particle in the 
high velocity tail, which entails less recoil events.
For simplicity we reduce the number of parameters in the velocity distribution
function to just two, fixing throughout $k=1$.

From Eqs.~\eqref{eqn:diff_rate} and \eqref{eqn:vel_distribution}, astrophysics affects the number of predicted recoil events through 
the dependence on the three velocities ($v_{\mbox{\tiny{Earth}}}$, $v_{\esc}$,
$v_d$) and the local density $\rho_{\loc}$. As done in Ref.~\cite{Pato:2010zk},
we note that $v_{\mbox{\tiny{Earth}}}$ can be reasonably approximated by the
value of the local circular velocity, $v_{\lsr}$, so we assume $v_{\mbox{\tiny{Earth}}} = v_{\lsr}$. The value of these ``astrophysical
nuisance parameters'' is constrained using a Gaussian prior with mean and standard deviation as motivated in Ref.~\cite{Pato:2010zk}, and given in the central part of Table~\ref{tab:nuis_params}.
In order to derive a prior for the velocity dispersion, $v_d$, we use the relationship
$v_d = \sqrt{3/2} v_0$ and, as in Ref.~\cite{Pato:2010zk}, the mean velocity $v_0$ is assumed to be equal
to $v_{\lsr}$. However, in our scans we allow both $v_d$ and 
$v_{\lsr}$ to vary as free, independent parameters (in order to be conservative), within the range given in 
Table~\ref{tab:nuis_params}.

Finally, $\sigmaSI$ is proportional to the square of the effective 
coupling of the neutralino to the proton, $f_p$:
\begin{equation}
\frac{f_p}{m_p} = \sum_{q=u,d,s} f_{Tq} \frac{\alpha_{q}^s}{m_q} +
\frac{2}{27} f_{TQ} \sum_{q=u,d,s} \frac{\alpha_{q}^s}{mq},
\end{equation}
where the quantities $f_{Tq}$ represent the contributions of the light quarks to the mass of the proton,
and are defined as
$f_{Tq}=m_p \langle p | m_q \bar{q}q | p \rangle$. The second term corresponds to the interaction of the neutralino with the gluon scalar density in the proton, with $f_{TQ}=1-\sum_{q=u,d,s}f_{Tq}$. 
The coefficients $\alpha_{q}^s$  correspond to the neutralino-quark scalar couplings in the low-energy effective Lagrangian and are calculated for each point in our scan. 
The hadronic matrix elements $f_{Tq}$ are determined experimentally in nuclear measurements 
and the uncertainty in their value (especially that on $f_{Ts}$) has a direct impact on the computed cross section. We therefore consider them as nuisance parameters in the scan 
and we constrain them with an informative Gaussian prior, with mean and standard deviation as indicated in Table~\ref{tab:nuis_params}, taken from~\cite{Ellis} (for an analysis of the capabilities of future ton-scale experiments to constrain dark matter including hadronic uncertainties, see~\cite{Akrami:2010dn}).

\subsection{Implementation of XENON100 data}
\label{secXenon100} 

We have now assembled all the necessary elements to implement the constraints from direct detection experiments to our likelihood.  
The most stringent results to date have been recently obtained by the XENON100 
experiment, which is located at the Laboratori Nazionali del Gran 
Sasso in Italy. DM is searched for with a target of dual-phase (liquid/gas)
xenon in an environment of extremely low background. A particle interacting
with the detector produces a primary scintillation signal $S1$ in the liquid 
xenon. Subsequently, ionization electrons drift towards the 
region with gaseous xenon producing the secondary scintillation signal $S2$.
The background is largely reduced by considering only single scattering events, while the $S2/S1$ ratio is employed to discriminate nuclear recoils 
(produced by backgrounds or candidate WIMPs event) from electronic recoils (from $\gamma$ 
or $\beta$ background).

The latest results~\cite{xenon:2011hi} have been obtained with an effective
volume of 48~kg and 100.9 days of exposure. The signal region is defined 
by a series of quality cuts determined blindly, in an energy window 
between 8.4 and 44.6~keV, from the requirement of having from 4 to 30 
photoelectron events (PE).
In this region, 3 events have been detected with energies of 12.1~keV, 30.2~keV 
and 34.6~keV. The expected number of events from the background is $b = 1.8 \pm 0.6$. In our analysis, we adopt a fixed central value for the background. We have checked that marginalizing over the background uncertainty has a negligible impact on our results.

In this work, we do not employ the energy distribution of the observed events, but only their total number. 
Having observed $\Nobs = 3$ events and with an
expected background of $b$ events, the likelihood function for the parameters $\params$ is a Poisson distribution (up to a normalization constant, $L_0$)
\begin{equation}
\mathcal{L}_\text{Xe100} (\params) \propto p(\Nobs|\lambda) = L_0 \frac{\lambda^{\Nobs}}{\Nobs  !} e^{-\lambda}\, ,
\label{eqn:poisson_likelihood}
\end{equation}
where $\lambda = s(\params)+b$ and the expected signal $s(\params)$ is obtained as described below. The normalization constant $L_0$ is chosen in such a way that $\ln {\mathcal L}_\text{Xe100} = 0$ when $\lambda = \hat{N}$, i.e., when the predicted signal plus the expected background match the observed number of counts. The probability distribution of the number of PE events, $n$, is 
obtained by convolving the differential recoil rate with a Poisson distribution with mean $S1(E)$, accounting for the acceptance cuts and other quality cuts of the data via a function $\zeta(E)$:
\be 
\frac{{\rm d} R }{{\rm d} n} = \int_0^\infty \frac{{\rm d} R }{{\rm d} E} \zeta(E) P(n|S1(E)) {\rm d} E\,,
\ee
where $\zeta(E)$ is a function parameterizing the acceptance of all data quality cuts, $P(n|S1(E))$ is a Poisson distribution for $n$ with mean $S1(E)$, and 
\be
S1(E) = L_y \frac{S_{\rm nr}}{S_{\rm ee}} E L_{\rm eff} (E)
\ee
gives the number of PE produced by an event with recoil energy $E$ (we neglect the uncertainty due to the finite single-electron resolution of the photomultipliers, which is small, $\sigma_{\rm PMT} = 0.5$ PE). In the above equation, $S_{\rm nr} = 0.95$ and $S_{\rm ee} = 0.58$ are the scintillation quenching factors for nuclear and electronic recoils, respectively; $L_y = 2.20 \pm 0.09$ PE/keV$_{\rm ee}$ is the light yield (whose uncertainty we neglect, as it is small) and $ L_{\rm eff} (E)$ is the scintillation efficiency, which we take to be the best fit line in Fig.~1 of~\cite{xenon:2011hi}. We neglect the uncertainty in $ L_{\rm eff} (E)$, which however is only important for very light WIMP mass, $m_\chi < 10$ GeV, as argued recently in Ref.~\cite{Arina:2011si}. As such small values of the neutralino mass are not realized in the cMSSM, our assumption has a negligible impact on our results. 
The expected total number of events is obtained by summing over all possible number of PE reaching the photomultipliers, within the acceptance window $\rm PE_{min} = 4$ and $\rm PE_{max} = 30$:
\be
s = \sum_{n = \rm PE_{min} }^{\rm PE_{max}}  \epsilon \frac{{\rm d} R }{{\rm d} n}, 
\ee
where $\epsilon$ is the exposure. We make the further simplification of assuming an energy-independent $\zeta$, and for the combined value of $\zeta \epsilon$ we take the effective (post-cuts) exposure of 1481 kg days~\cite{xenon:2011hi}.

We have checked that the above simplified treatment reproduces fairly accurately the exclusion limit obtained in Ref.~\cite{xenon:2011hi} in the $m_\chi-\sigmaSI$  plane. As we do not make use of spectral information, we expect our limit to be more conservative than the one derived using the more sophisticated likelihood in Ref.~\cite{xenon:2011hi,Aprile:2011hx}. Indeed, for the same choice of astrophysical nuisance parameters as in Ref.~\cite{xenon:2011hi}, we find a 90\% limit $\sigmaSI = 0.85 \times 10^{-8}$ pb for a WIMP mass $m_\chi = 50$ GeV, which is to be compared with the tighter limit found in Ref.~\cite{xenon:2011hi} , $\sigmaSI = 0.70 \times 10^{-8}$ pb.

\section{Results}
\label{secresults}

\begin{figure*}
\centering
\includegraphics[width=0.32\linewidth]{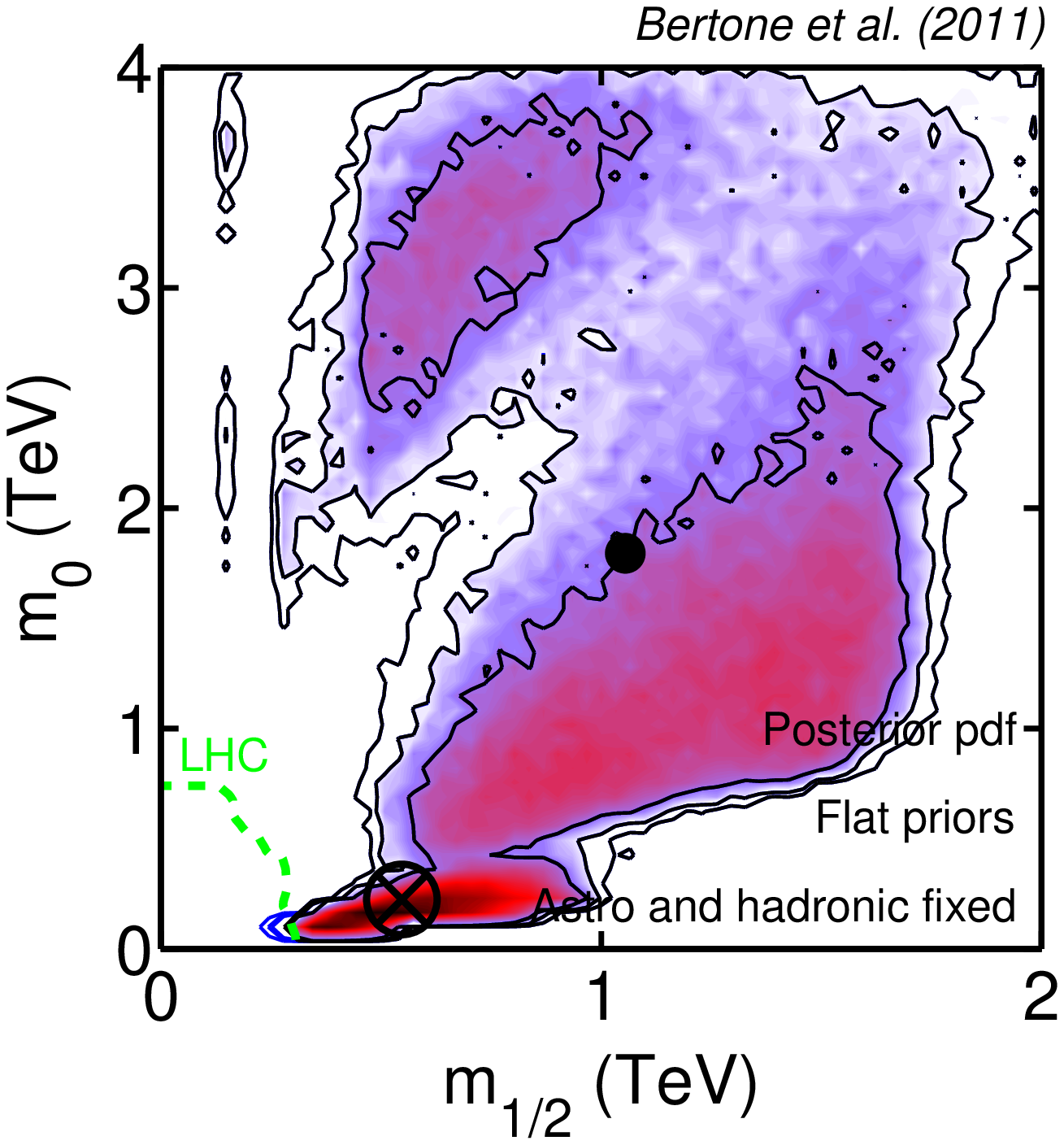}
\includegraphics[width=0.32\linewidth]{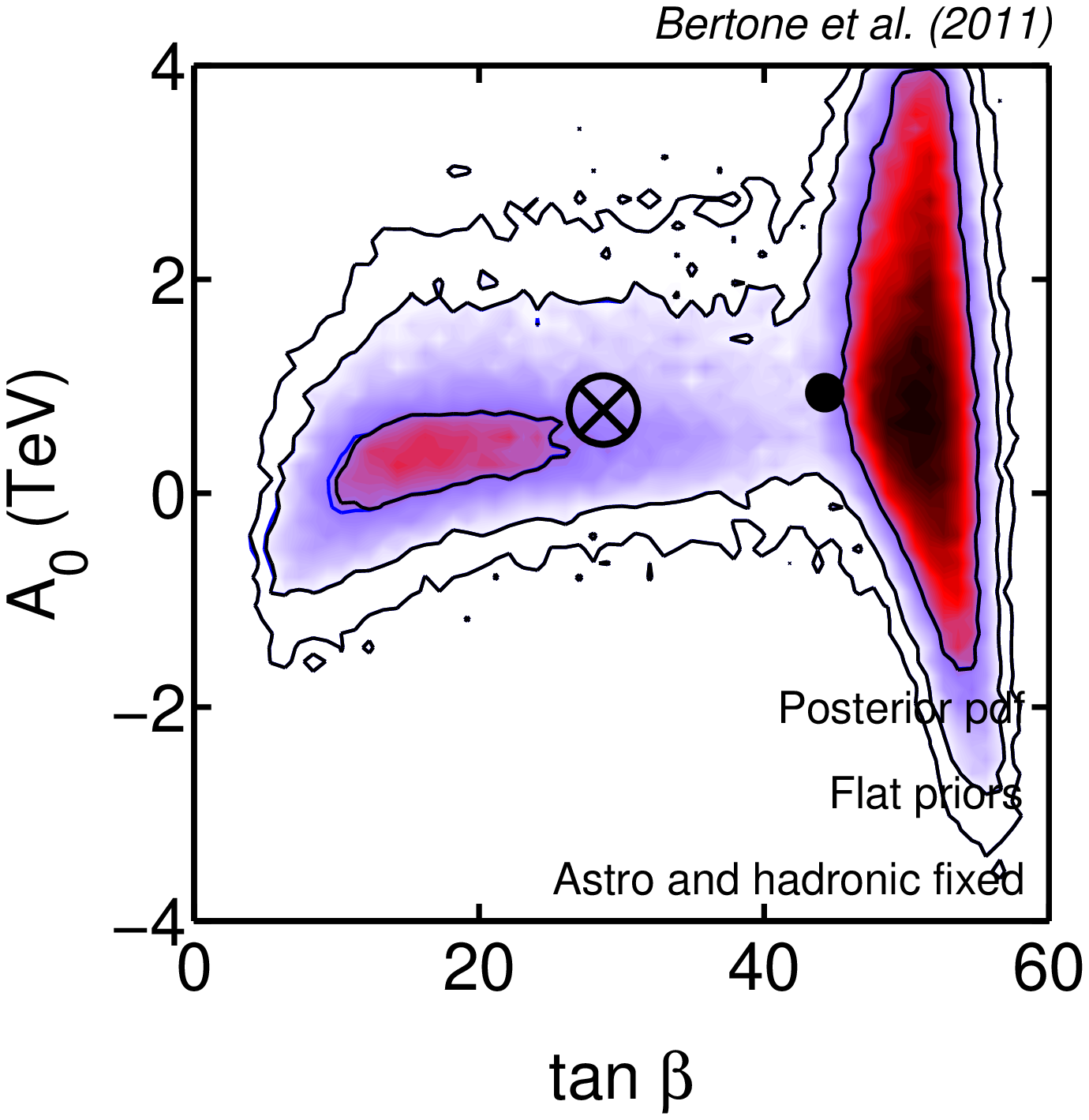}
\includegraphics[width=0.32\linewidth]{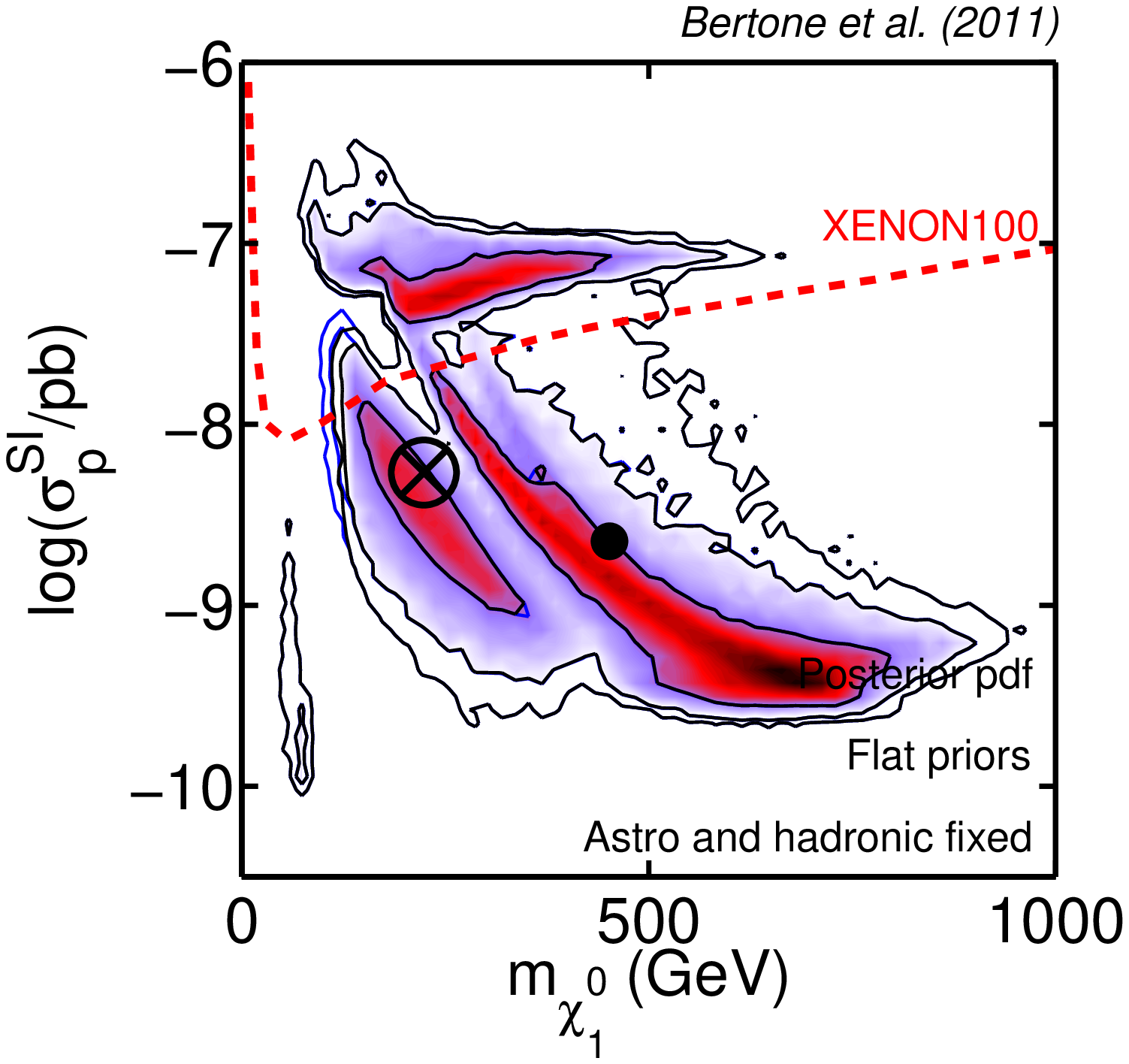} \\
\includegraphics[width=0.32\linewidth]{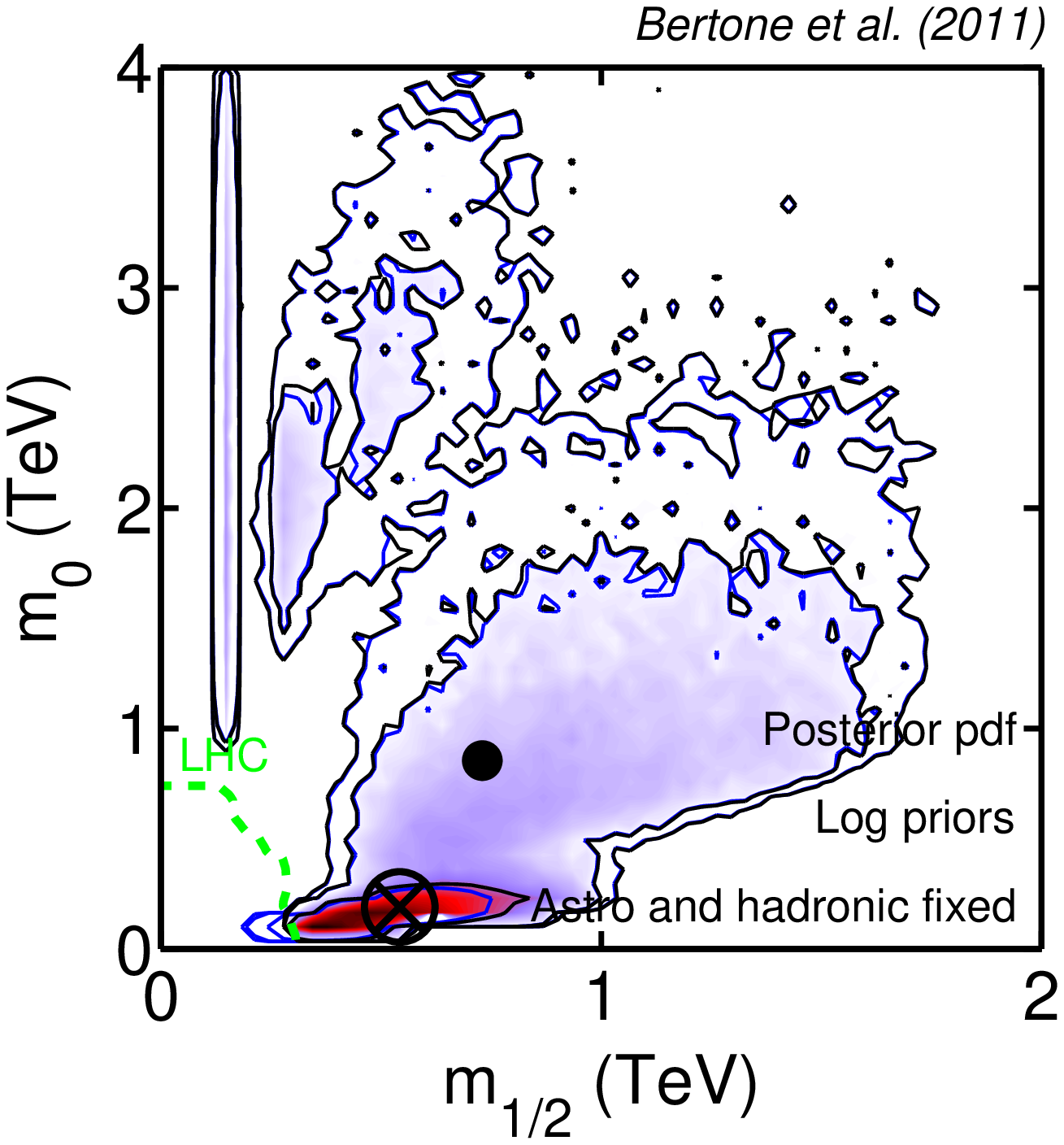}
\includegraphics[width=0.32\linewidth]{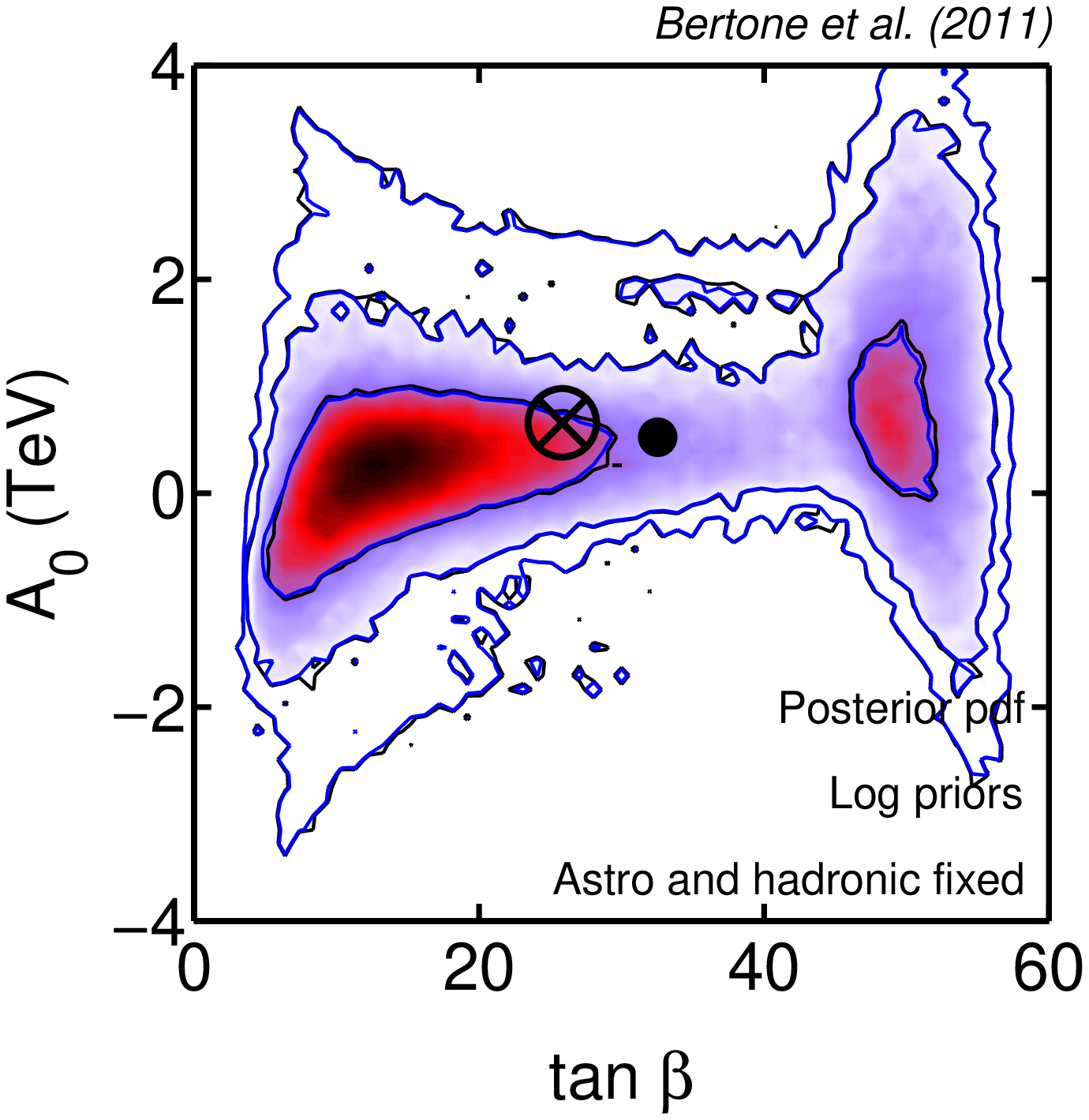}
\includegraphics[width=0.32\linewidth]{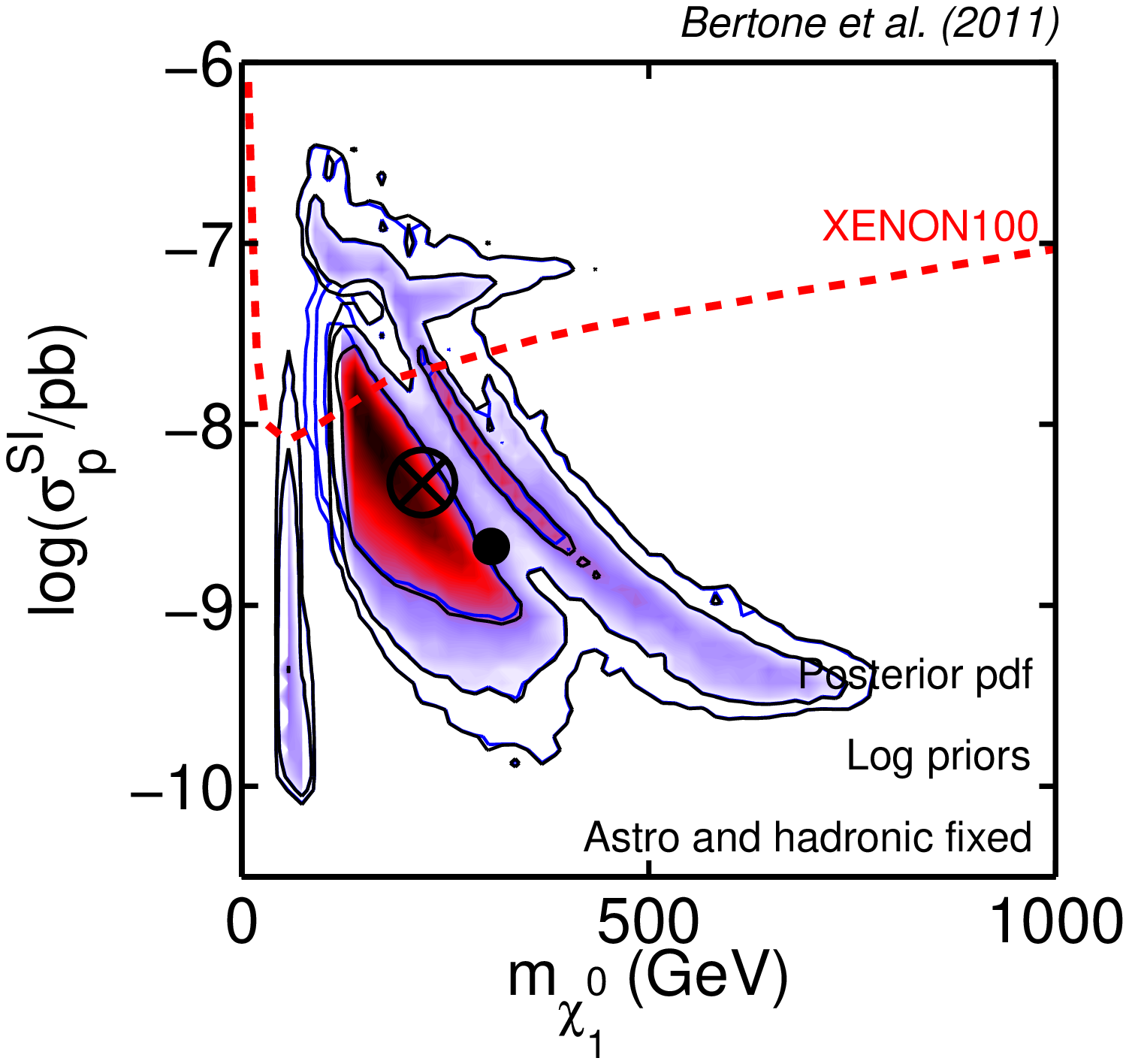}
\includegraphics[width=0.32\linewidth]{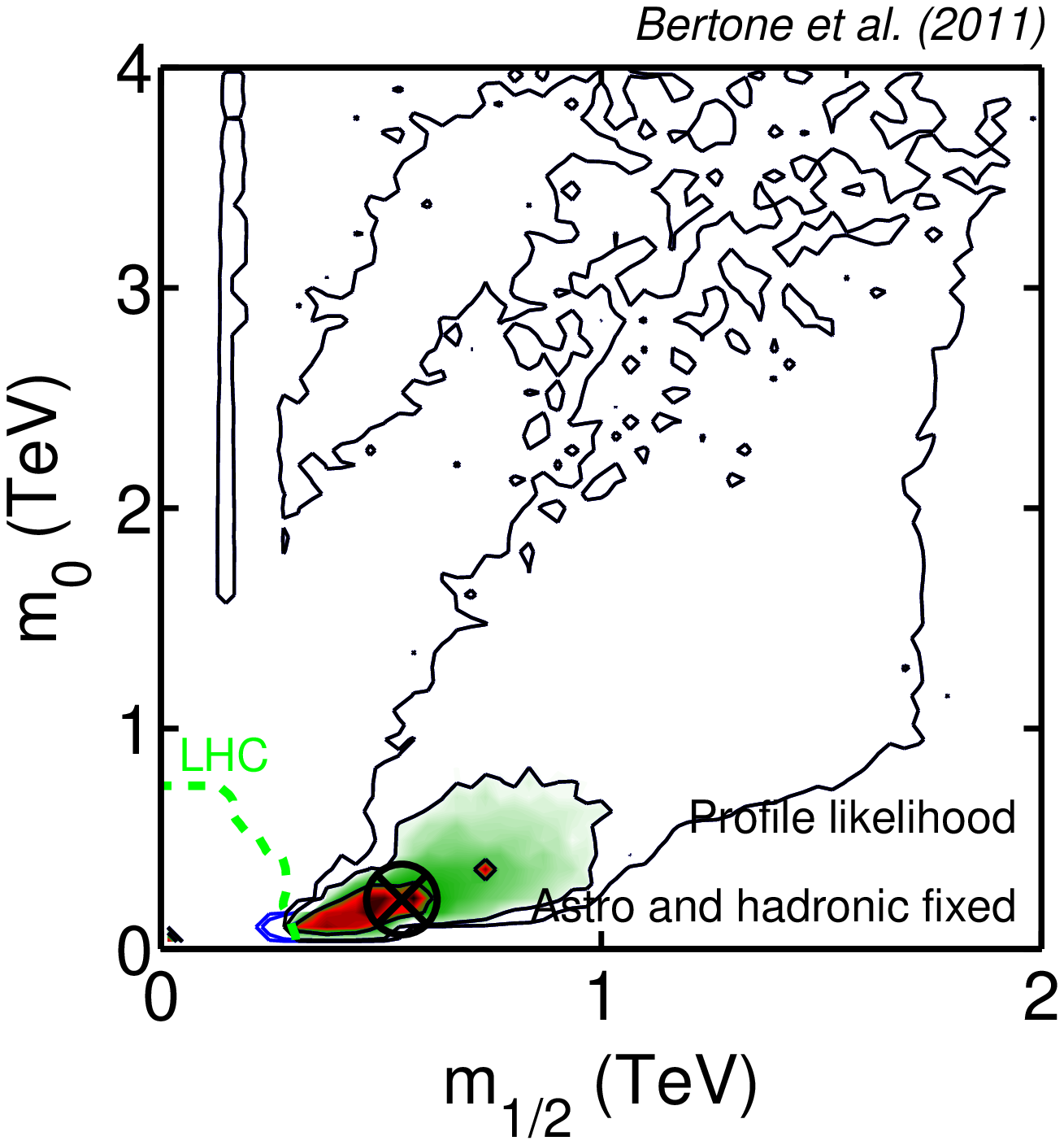}
\includegraphics[width=0.32\linewidth]{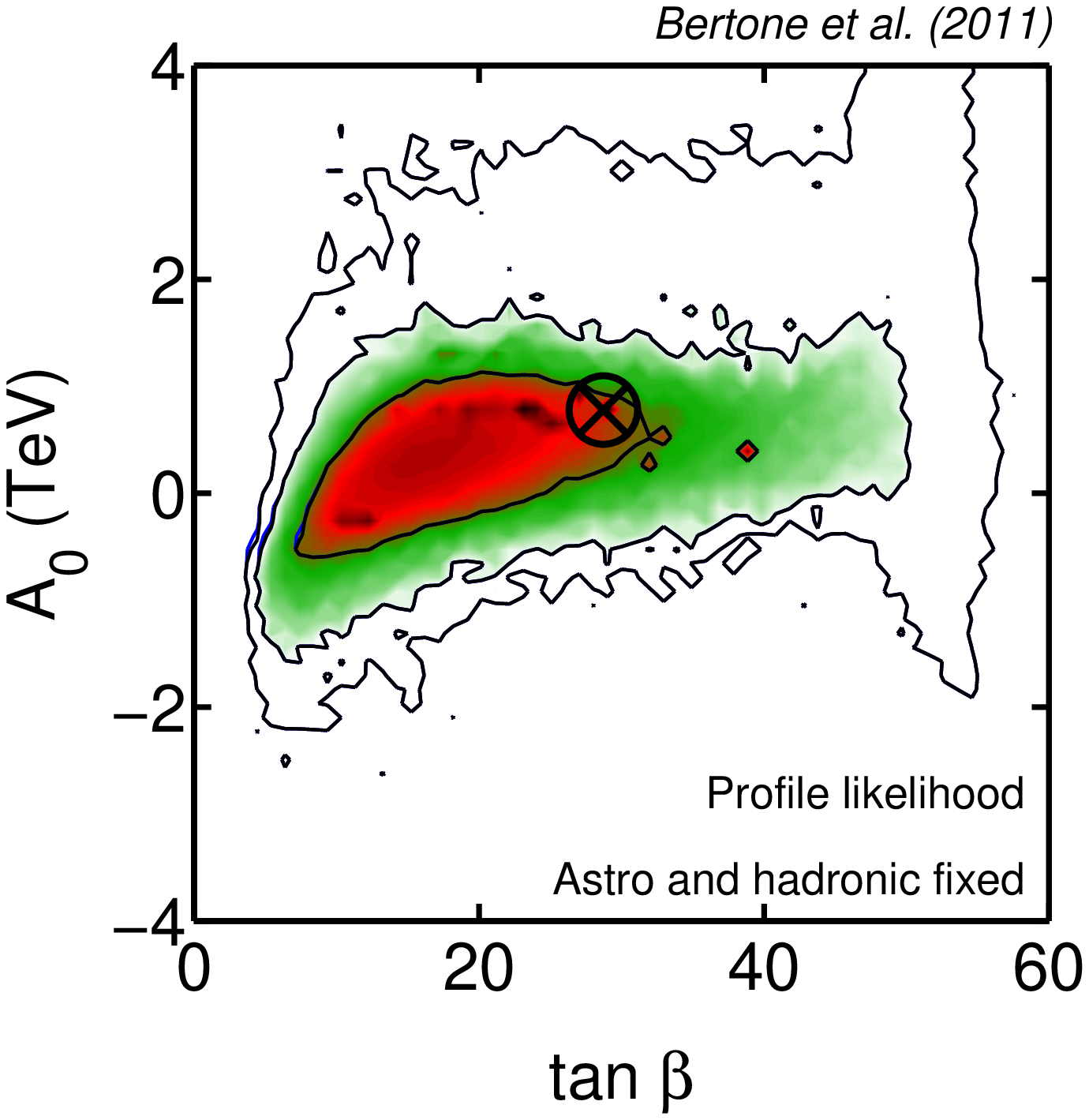}
\includegraphics[width=0.32\linewidth]{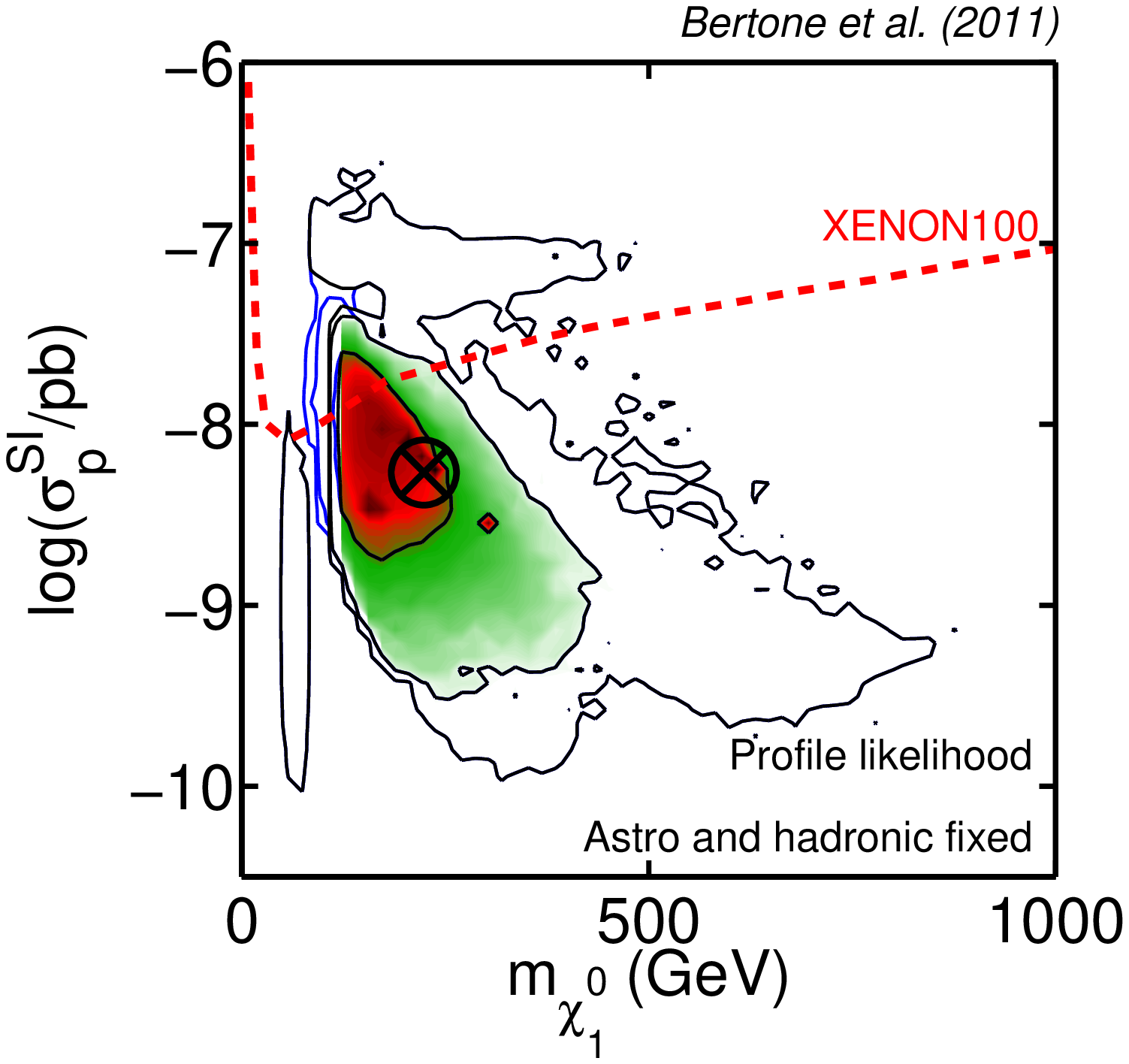}
\caption{\fontsize{9}{9} \selectfont Black contours: posterior pdf (upper panels, for flat and log priors) and profile likelihood (lower panels) for the cMSSM parameters, including all present-day constraints (WMAP 7-years and LHC first results included), except XENON100. From the inside out, contours enclose 68\%, 95\% and 99\% of  marginal posterior probability (top two rows) and the corresponding profiled confidence intervals (bottom panels). The black cross represents the best fit, the black dot the posterior mean (for the pdf plots). Parameters describing astrophysical and hadronic uncertainties have been fixed to their fiducial values. Blue contours represent the constraints obtained without inclusion of LHC data. In the plots on the left, the dashed/green line represents the current LHC exclusion limit, while in the right-most plots the red/dashed line is the 90\% exclusion limit from XENON100, from Ref.~\cite{xenon:2011hi}, rescaled to our fiducial local DM density of $\rho_{\loc} = 0.4$ GeV/cm$^3$. \label{fig:ALL_NoXe_LHC}}  
\end{figure*}

\begin{table*}
\begin{center}
\begin{tabular}{|l|ll| l l|}
\hline
& \multicolumn{2}{c}{Fixed nuisance} &  \multicolumn{2}{c|}{Incl. nuisance}\\
& Flat priors & Log priors & Flat priors & Log priors \\ \hline
\multicolumn{5}{|c|}{Input parameters}\\\hline
$ m_0$ [GeV] & 108.52 & 88.68 & 220.51 & 98.26 \\
$ m_{1/2}$ [GeV] & 390.96 & 370.54 & 582.41 & 363.28 \\
$A_0$ [GeV] & -326.06 & -286.55 & 990.39 & -278.95 \\
$ \tan\beta$ & 13.94 & 11.10 & 24.29 & 13.48 \\
\hline 
$M_t$ [GeV] & 173.383 & 173.129 & 173.209 & 173.617 \\
$m_b(m_b)^{\bar{MS}}$ [GeV] & 4.210 & 4.236 & 4.212 & 4.271 \\
$[\alpha_{em}(M_Z)^{\bar{MS}}]^{-1}$ & 127.956 & 127.968 & 127.959 & 127.961 \\
$\alpha_s(M_Z)^{\bar{MS}}$ & 0.118 & 0.119 & 0.117 & 0.117 \\ \hline
\multicolumn{5}{|c|}{Observables}\\\hline
$m_h$ [GeV] & 115.1 & 114.2 & 114.7 & 114.5 \\
$m_\chi$ [GeV] & 157.6  & 148.5 & 238.8 & 145.6 \\
$\delta a_\mu^{SUSY} \times 10^{10}$ & 18.54 & 17.22 & 29.20 & 20.89 \\ 
$BR(\bar{B} \rightarrow X_s\gamma) \times 10^4$ & 3.83 & 3.36 & 3.16 & 3.59 \\
$\DeltaO  \times 10^{2}$ & 8.78 & 8.42 & 8.25 & 8.45 \\ 
$\Dstaunu \times 10^{2}$ & 4.82 & 4.82 & 4.82 & 4.82 \\
$\Dmunu \times 10^{4}$ & 3.86 & 3.86 & 3.86 & 3.86 \\
$\sigmaSI$ [pb] &  $3.2 \times 10^{-9}$ & $3.5 \times 10^{-9}$ & $4.0 \times 10^{-9}$ & $3.5 \times 10^{-9}$ \\
$\sigmaSD$ [pb] & $8.0 \times 10^{-7}$  & $1.1 \times 10^{-6}$ & $7.9 \times 10^{-7}$ & $1.1 \times 10^{-6}$ \\
$\Omega_\chi h^2$ & 0.1142 & 0.1142 & 0.1096 & 0.1105 \\ 
\hline
\end{tabular}
\end{center}
\caption{Best fit points for the flat and log prior scans. The columns ``Fixed nuisance'' are for the case where astrophysical and hadronic parameters have been fixed to their fiducial values, while ``Incl. nuisance'' denotes the case where they have been included in the scan and profiled over. \label{tab:bestfit}}
\end{table*}

\begin{table*}
\begin{center}
\begin{tabular}{|l|ll| l l|}
\hline
& \multicolumn{2}{c}{Fixed nuisance} &  \multicolumn{2}{c|}{Incl. nuisance}\\
Observable &   Flat priors & Log priors  & Flat priors & Log priors \\ \hline
SM nuisance & 0.204 & 0.750 & 0.087 & 1.247 \\
Astro nuisance &  N/A & N/A & 1.060 & 0.522 \\
Hadronic nuisance & N/A & N/A & 0.630 & 0.667 \\
$M_W$ & 0.986 & 1.072 & 1.037 & 0.828 \\
$\sin^2\theta_{eff}$ & 0.105 & 0.900 & 0.096 & 0.150 \\
$\delta a_\mu^{SUSY} $ & 1.755 & 2.202 & 0.002 & 1.089 \\
$BR(\bar{B} \rightarrow X_s\gamma) $ & 0.527 & 0.228 & 0.964 & 0.008 \\
$\Delta M_{B_s}$ & 0.274 &  0.262 & 0.276 & 0.315 \\
$\RBtaunu$  &  0.647 & 0.614 & 0.780 & 0.656 \\
$\DeltaO  \times 10^2$  & 3.816 & 3.306 & 3.079 & 3.356 \\
$\RBDtaunuBDenu $ & 0.818 & 0.802 & 0.843 & 0.820 \\
$\Rl$ & 0.345 & 0.340 & 0.367 & 0.347 \\
$\Dstaunu $ & 2.224 & 2.220 & 2.238 & 2.225 \\
$\Dsmunu  $ & 3.109 & 3.106 & 3.122 & 3.109 \\
$\Dmunu $  & 0.008  & 0.084 & 0.008 & 0.008 \\
$\Omega_\chi h^2$ & 0.025 & 0.026 & 0.056 & 0.022 \\ \hline
$m_h$ & 1.022 & 1.482 & 1.216 & 1.297 \\
XENON100 & 0.042 & 0.110 & 0.0 & 0.165 \\
LHC & 0.0 & 0.0 & 0.0 & 0.0 \\ 
Sparticles & 0.0 & 0.0 & 0.0 & 0.0 \\ 
$\brbsmumu$ & 0.0 & 0.0 & 0.0 & 0.0 \\ \hline
Total & 15.91 & 16.62 & 15.86 & 16.83 \\
$\chi^2/$dof & 1.77 & 1.85 & 1.76 & 1.87  \\
\hline
\end{tabular}
\end{center}
\caption{Breakdown of the total $\chi^2$ by observable for the best fit points in our scans, for both cases where the nuisance parameters have been fixed (two left columns) and profiled over (two right columns). \label{tab:chisquare} }
\end{table*}

\subsection{Impact of LHC data}

We start by showing in Fig.~\ref{fig:ALL_NoXe_LHC} the impact of the new LHC results on the posterior pdf (for both choices of priors) and the profile likelihood of the cMSSM. As one can see by comparing the empty grey contours (without LHC data) with the black contours (including LHC limits), the impact of current LHC limits is actually fairly modest, since the largest part of the favoured region under the remaining data is located at larger values of $m_{0}$ and $m_{1/2}$ (see also Refs.~\cite{Allanach:2011wi,Buchmueller:2011ki} for a more detailed analysis).

The posterior pdf's in this figure exhibit the familiar features of the cMSSM (see e.g.~the discussion in Ref.~\cite{Trotta:2008bp}).
 We observe in particular two distinct areas in the $m_0, m_{1/2}$ plane: the neutralino-stau
coannihilation region, that appears as a narrow band for small scalar
masses, and a wide area which extends to sizable scalar masses, corresponding to the so-called focus point (FP) region \cite{Feng:1999zg,Feng:2000gh} 
(also know as hyperbolic branch \cite{Chan:1997bi}). The latter is particularly prominent in the case of flat priors (top panels), due to the ``volume effects'' associated with this prior discussed in Ref.~\cite{Trotta:2008bp}. Along the FP branch the absolute value of the $\mu$ parameter is of order of the
electroweak scale
despite scalar particles  
being very massive (generally several TeV), as long as gaugino masses
are not too large. The smallness of $\mu$ 
has been interpreted
as corresponding to low values of the fine-tuning of the model and is
due to a ``focused'' behaviour of the Renormalization Group Equation
for the Higgs mass parameters.

This FP region has
interesting consequences for the nature of the dark matter. Since
$\mu$ is of the same order as the gaugino mass parameters, the
neutralino exhibits a significant Higgsino fraction (it is a mixed
Higgsino-bino state). As a
consequence its relic density decreases \cite{Baer:1997ai}, making it
possible to satisfy the experimental constraint on the abundance of
dark matter.
The increased Higgsino fraction also leads to a large
spin-independent neutralino-nucleon cross section, since Higgs
exchange along a $t$-channel becomes more effective (it is proportional to the product of Higgsino and gaugino components). 
The cross section can thus 
be as large as $\sigmaSI\sim 10^{-8}$ pb for neutralino masses as large as $1$
TeV, and in fact the focus point region is easily identified in the $m_\chi\-\sigmaSI$ plane (see the top right panel of Fig.~\ref{fig:ALL_NoXe_LHC}) as a plateau in the region $\sigmaSI \sim 10^{-7} - 10^{-8}$ pb. 

As we shall see, the latest XENON100 direct detection data set stringent
constraints on this region. This can already be seen from the right most panels in Fig.~\ref{fig:ALL_NoXe_LHC}, which display the constraints (without inclusion of direct detection data) in the plane spanned by $m_\chi$ and $\sigmaSI$: especially for the flat prior case, the FP region appears as a sizable island of high probability above the dashed red line, which represents the 90\% limit from the most recent XENON00 data (for standard astrophysical assumptions -- we shall relax those in the next section). Therefore we expect that the inclusion of the direct detection constraints will further reduce the viability of the FP region, as we demonstrate below. 

Turning now to the profile likelihood results (bottom panels in Fig.~\ref{fig:ALL_NoXe_LHC}), we see that the FP region is excluded at the $95\%$ level, but still viable at the 99\% level (outer-most black contours), which we are able to map out thanks to our high-resolution scan. Most of the favoured region however lies below $\sim 1$ TeV for both the scalar and gaugino masses. As a consequence, we observe in the $m_\chi - \sigmaSI$ (bottom right panel) that the cMSSM parameter above the current nominal XENON100 limit is for the most part disfavoured at the 99\% level from a profile likelihood perspective. We also point out that the 99\% region from the profile likelihood is much wider than could be assumed just by qualitatively extending either the 68\% or the 95\% range, and this owing to the highly non-Gaussian nature of the tails of the distribution. Our results therefore indicate that a high-resolution scan is necessary to map out the tails of the profile likelihood with sufficient accuracy in order to delimit the 99\% region, whose extent is much larger than would be inferred by assuming an approximately Gaussian distribution from the 68\% region. Finally, it is interesting that the extent of the 99\% profile likelihood region is actually very similar to the 99\% posterior region from the flat prior scan (compare outer-most contours of the middle and bottom panels of Fig.~\ref{fig:ALL_NoXe_LHC}).

We now turn to discuss the best fit points identified in our scans (see Table~\ref{tab:bestfit}). The contribution of each observable to the best fit $\chi^2$ is shown in Table~\ref{tab:chisquare}, where the $\chi^2$ is defined as
\begin{equation}
\chi^2 \equiv -2 \sum_{i} \ln {\mathcal L}_i ,
\end{equation}
and ${\mathcal L}_i $ is the likelihood for each observable.
The likelihood ${\mathcal L}_i$ is normalized in such a way that  for Gaussian-distributed data points $\ln {\mathcal L}_i = 0$ when the theoretical value matches the observed mean in Table~\ref{tab:SM_constraints}. For observables for which only limits exist, the likelihood is normalized so that $\ln {\mathcal L}_i = 0$ asymptotically above/below the stated exclusion limit. The one exception to this is the XENON100 likelihood, which is normalized as explained above, and therefore there is a residual contribution to the $\chi^2$ of $\approx 0.57$ units when $s(\params) \rightarrow 0$ (i.e. for very small interaction cross-sections, $\sigmaSI \ll 10^{-9}$ pb). In evaluating the number of degrees of freedom (dof), we only count as ``active'' Gaussian data points (13 in total, from Table~\ref{tab:SM_constraints}), for 4 cMSSM free parameters (nuisance parameters are not counted as each one of them is independently constrained), thus giving dof = 9.
Despite the leakage of probability into the FP discussed above, it is interesting that the best fit points for both the flat and log prior scans are very well compatible, and lie in the bulk region, in contrast to what had been found in Ref.~\cite{Akrami:2009hp,Feroz:2011bj}, who reported best fits in the FP region based on an older combination of data. Despite this, as discussed above the FP region cannot be excluded at the 99\% level either from a Bayesian or a profile likelihood perspective unless one includes in the likelihood direct detection data (see below).

The observables contributing the most to the total $\chi^2$ in Table~\ref{tab:chisquare}, namely 
the branching ratios of $\Dstaunu$ and $\Dsmunu$
are sensitive to new physics mainly through the mass of the charged 
Higgs bosons $H^\pm$ and to some extend to $\tan\beta$~\cite{Akeroyd:2009tn}, in particular for low $m_{H^\pm}$ and large $\tan\beta$ values. 
Our best fit points have relatively large 
$m_{H^\pm}$ and low/intermediate $\tan\beta$, and therefore the theoretical values for those processes are 
SM-like and have a discrepancy at the $1.5 \, \sigma$ level with the 
experimental values.
For the isospin asymmetry $\DeltaO$ new physics contributions 
become important to large $\tan\beta$ and low $m_{1/2}$~\cite{Ahmady:2006yr}. 
Hence the predictions for our best fit points are again SM-like, showing a discrepancy at the 
$2 \, \sigma$ level with the experimental central values. In constrast, relatively good fits can be found in the bulk region for both the anomalous magnetic moment of the muon and $\brbsgamma$, as can be seen from their relatively low contribution to the total $\chi^2$.

\begin{figure*}
\centering
\includegraphics[width=0.32\linewidth]{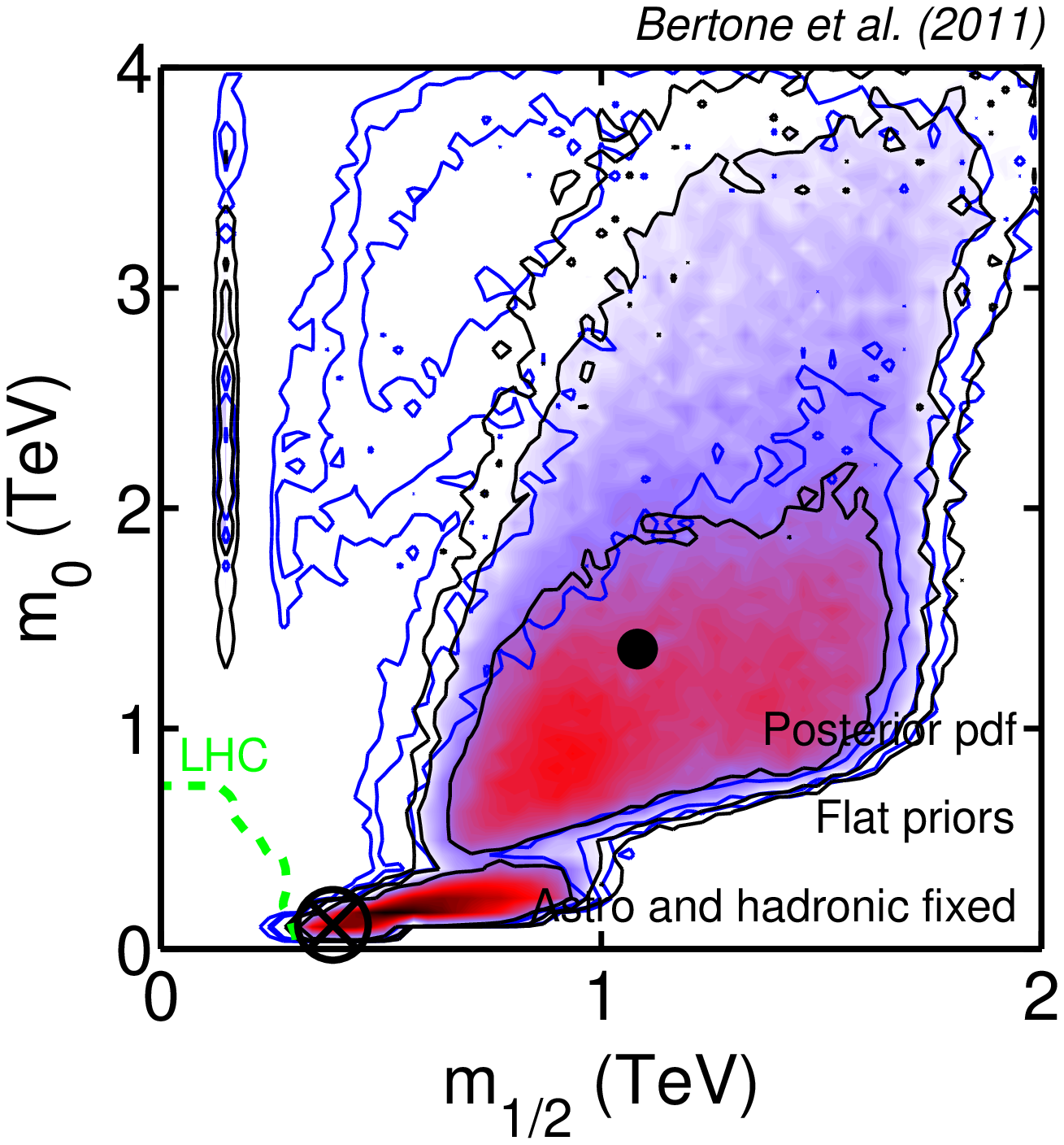}
\includegraphics[width=0.32\linewidth]{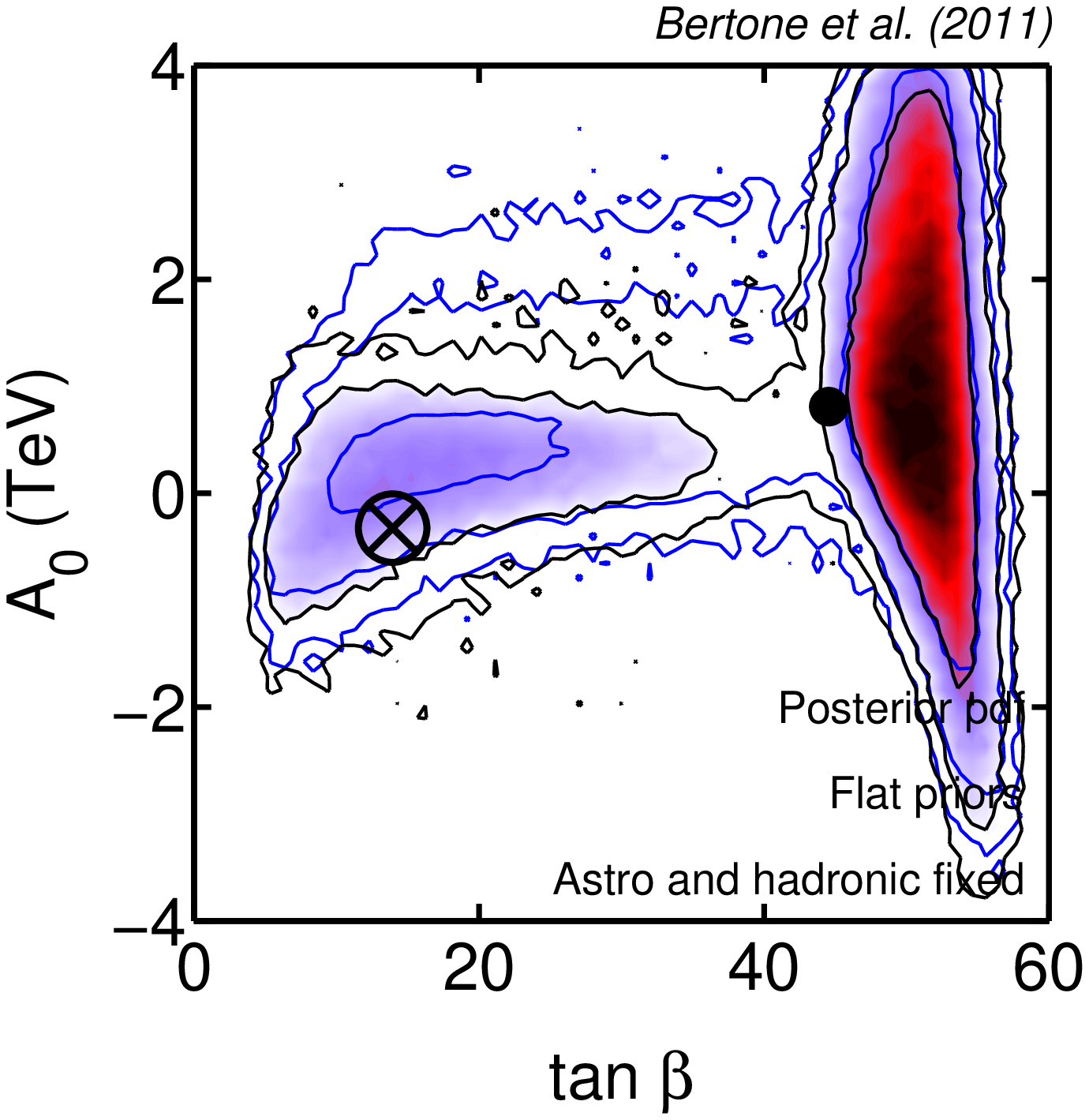}
\includegraphics[width=0.32\linewidth]{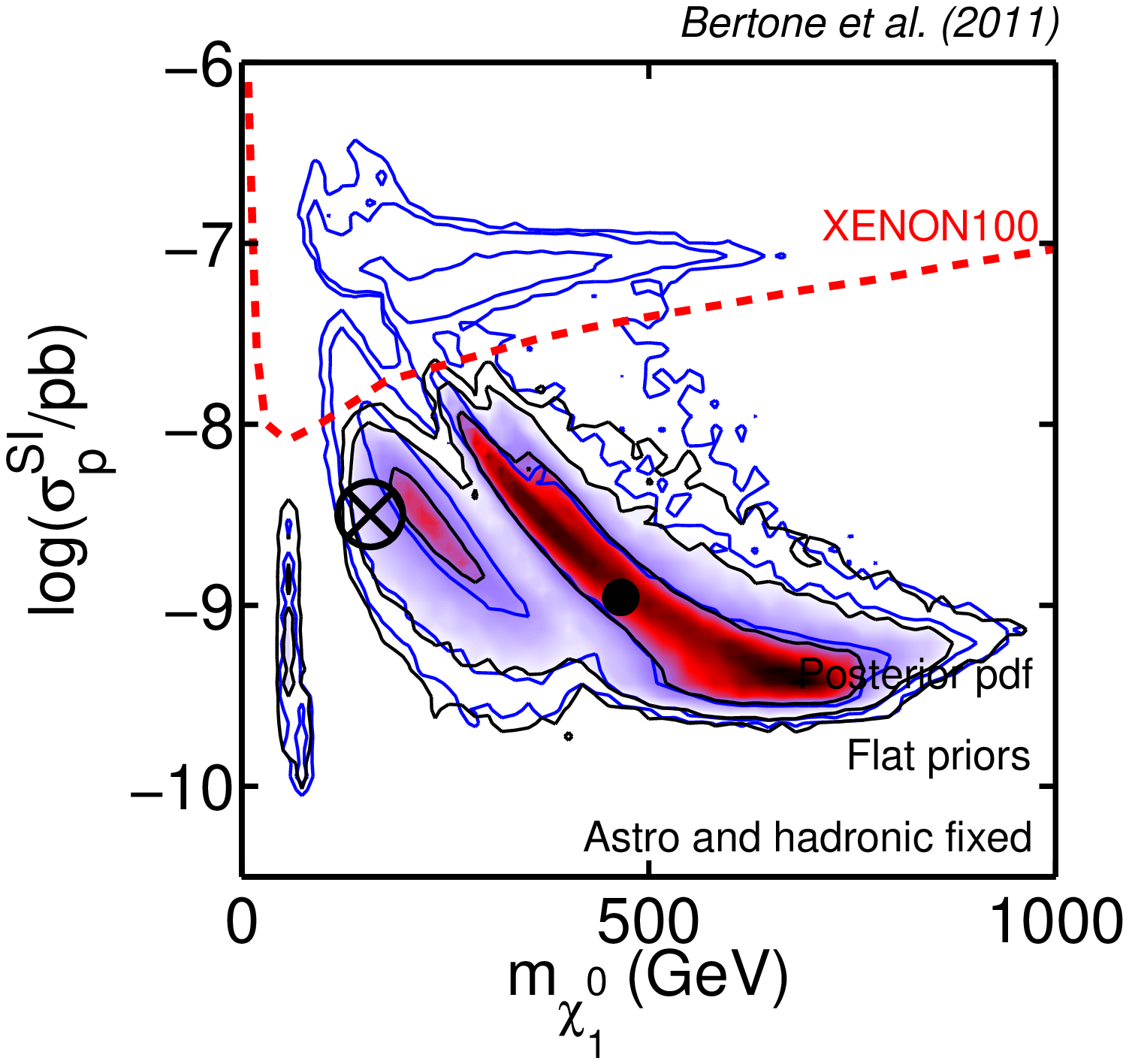}
\includegraphics[width=0.32\linewidth]{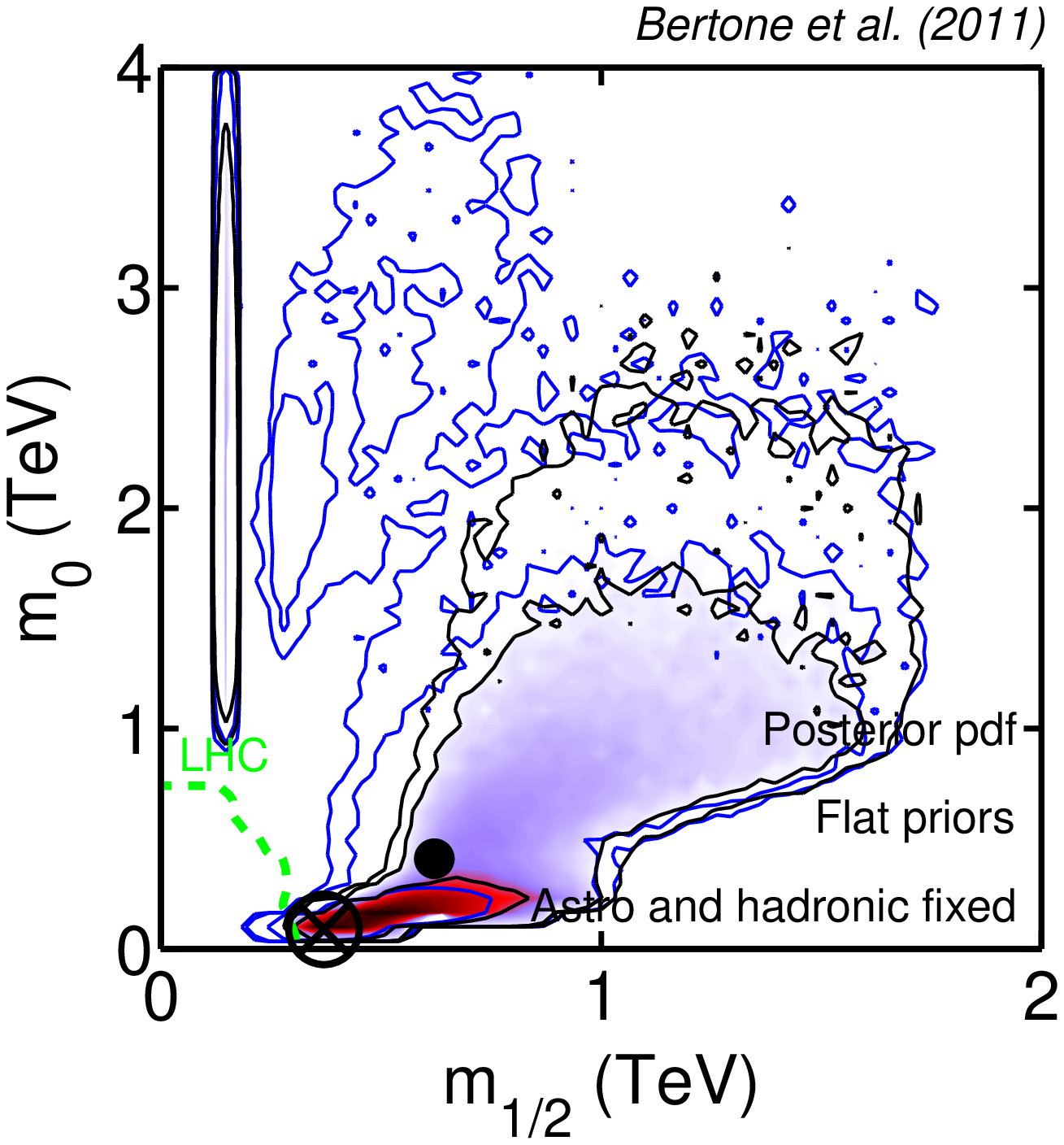}
\includegraphics[width=0.32\linewidth]{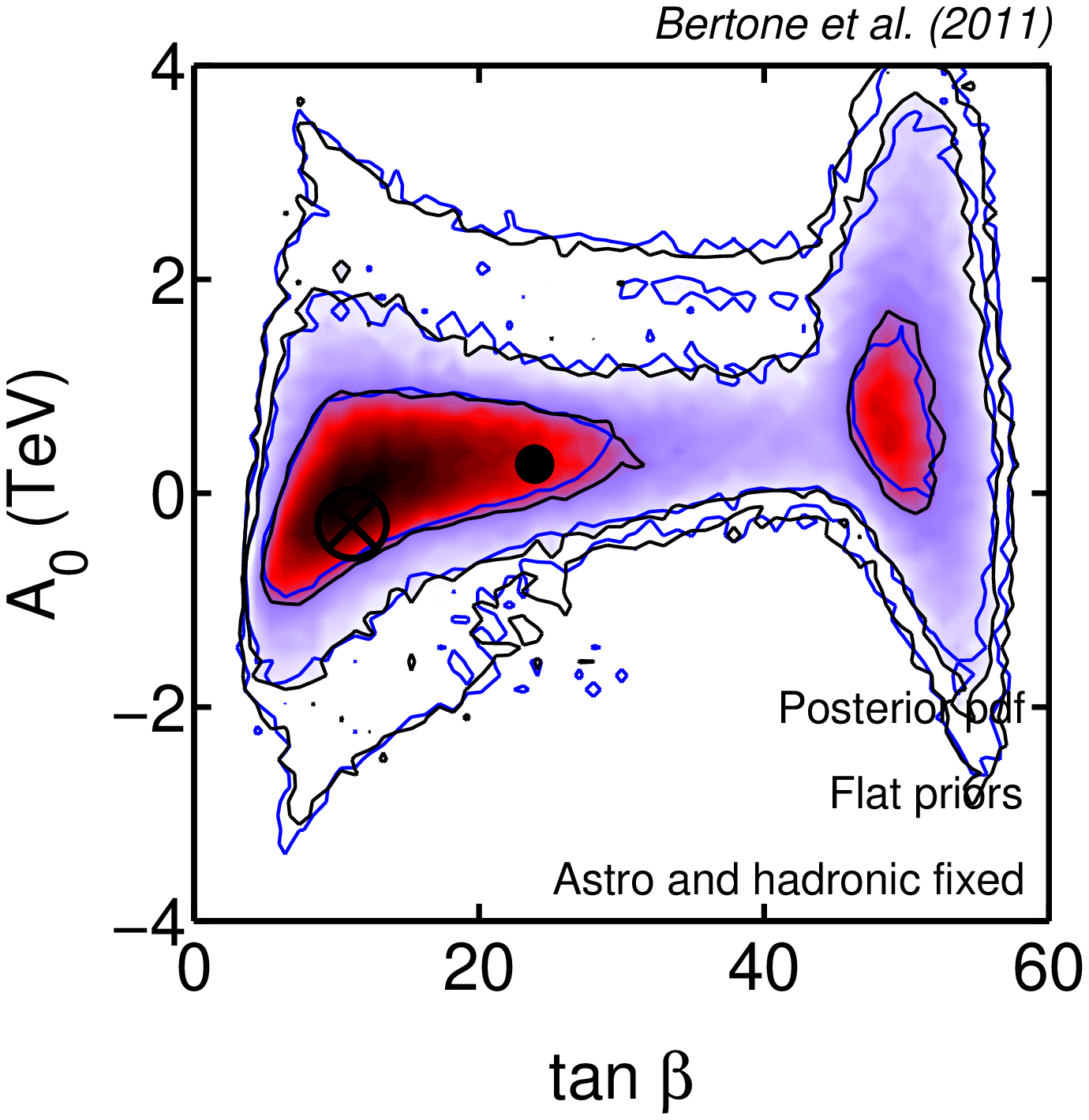}
\includegraphics[width=0.32\linewidth]{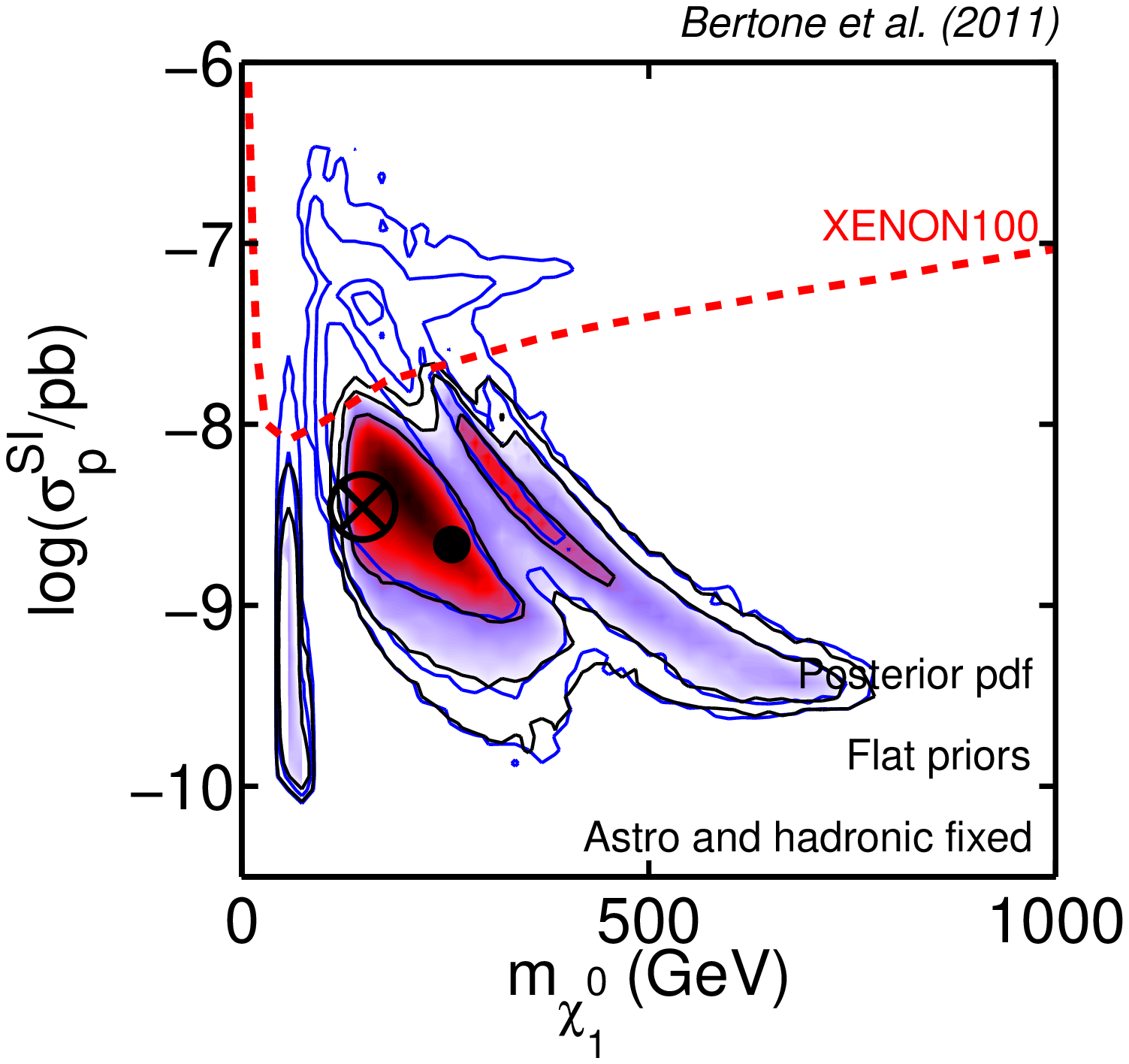}
\includegraphics[width=0.32\linewidth]{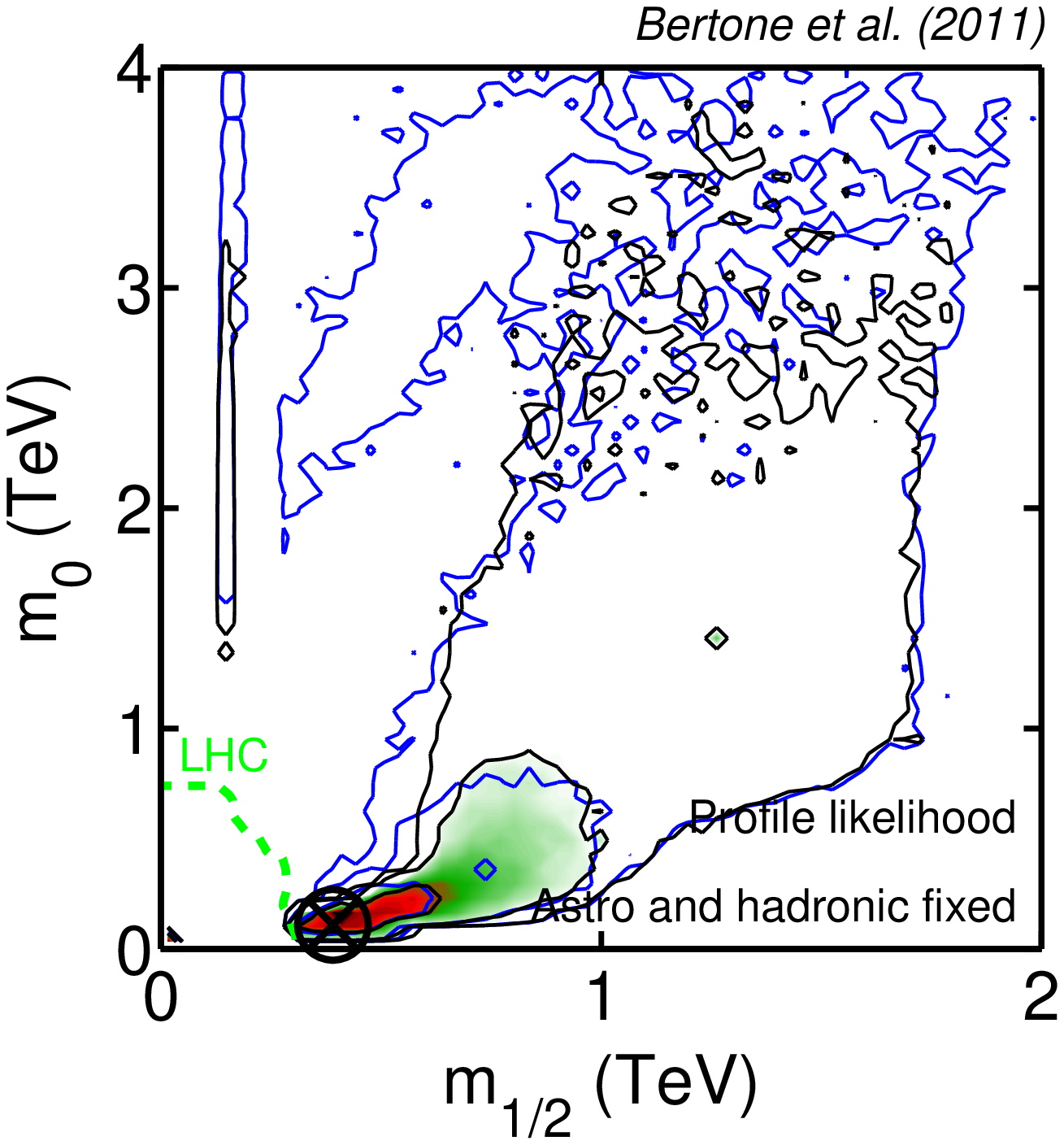}
\includegraphics[width=0.32\linewidth]{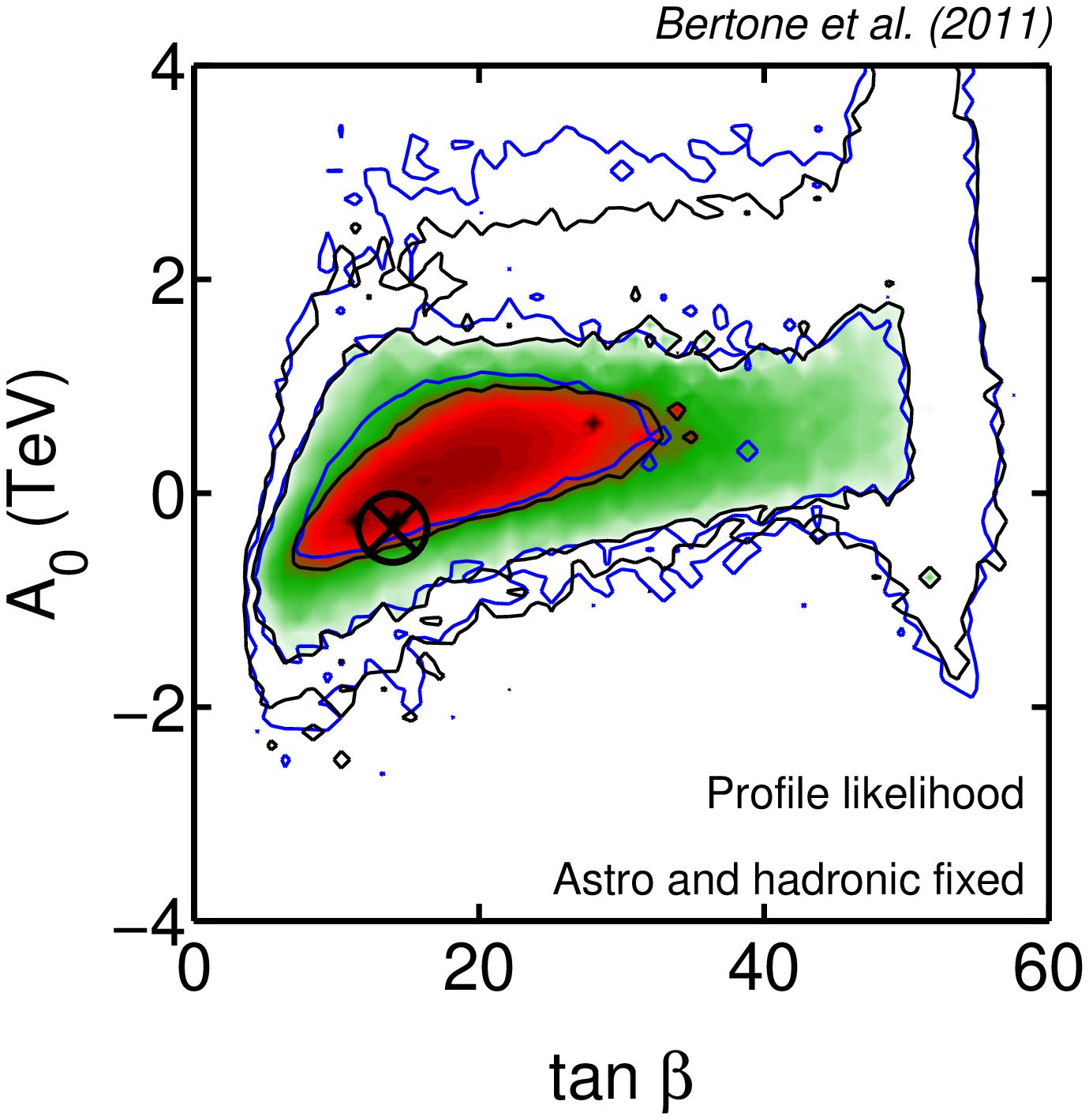}
\includegraphics[width=0.32\linewidth]{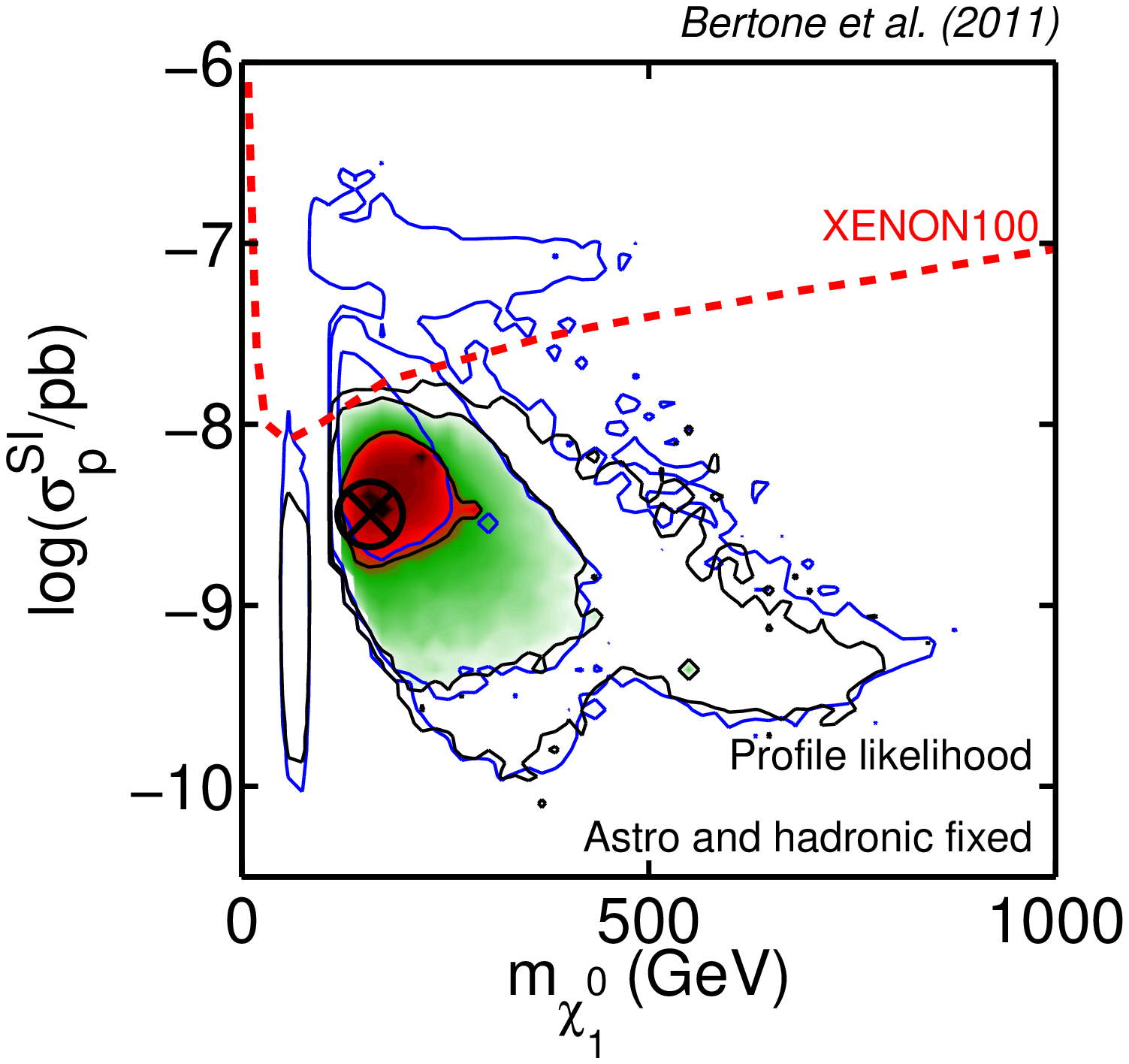}
\caption{\fontsize{9}{9} \selectfont As in Fig.~\ref{fig:ALL_NoXe_LHC}, but now black contours include XENON100 data (for fixed hadronic and astrophysical uncertainties), while the blue empty contours show for comparison the case where no direct detection data are included (from the inside out: 68\%, 95\% and 99\% regions). We observe a strong suppression of the viability of the FP region in the Bayesian posterior (top and middle row), and a better agreement between the posterior distributions and the profile likelihood. Notice that the XENON100 90\% limit (red/dashed line) has been included only to guide the eye, as our implementation of the XENON100 data is slightly more conservative than the procedure adopted in Ref.~\cite{xenon:2011hi}. \label{fig:ALL_Xe_LHC}}
\end{figure*}

\subsection{Impact of XENON100 for fixed nuisance parameters}

We show in Fig.~\ref{fig:ALL_Xe_LHC} the impact of further adding XENON100 data, and compare the resulting constraints with the case where all constraints are included except XENON100 (gray contours). We keep for the moment all astrophysical and hadronic nuisance parameters fixed to their fiducial values, before moving on to the full analysis. The impact of direct detection data on the posterior pdf with flat priors is fairly dramatic, as can be seen in the three panels of the first row. The most obvious consequence is that the FP becomes strongly disfavoured. In the case of log priors (second row of Fig.~\ref{fig:ALL_Xe_LHC}), the impact of direct detection data is less dramatic on the posterior, but the parameter space shrinks significantly also in this case, a result that clearly demonstrates the potential of direct detection experiments. Overall, we observe that the posterior pdf's for the two choices of priors are now in much better agreement with each other (despite some residual volume effects, which are evident e.g. in the large concentration of probability density for large $\tan\beta$ in the middle panel of the first row), as one would expect when the posterior becomes more and more strongly dominated by the likelihood and the impact of priors is reduced (see Ref.~\cite{Roszkowski:2009ye} for another example). The profile likelihood 99\% region (bottom panels) is also strongly reduced by inclusion of XENON100 data, and it now extends somewhat in-between the 99\% posterior region from the flat and the log prior scans. 

Regarding prospects for DM direct detection in the cMSSM, by comparing the right-most panels in Fig.~\ref{fig:ALL_Xe_LHC} we observed that the extension of the favoured region at 99\% in the $m_\chi, \sigmaSI$ plane is now almost independent of the statistical approach and the priors used, a point we will return to in more detail below when we discuss prospects for cMSSM discovery in various observational channels.

\begin{figure*}
\centering
\includegraphics[width=0.32\linewidth]{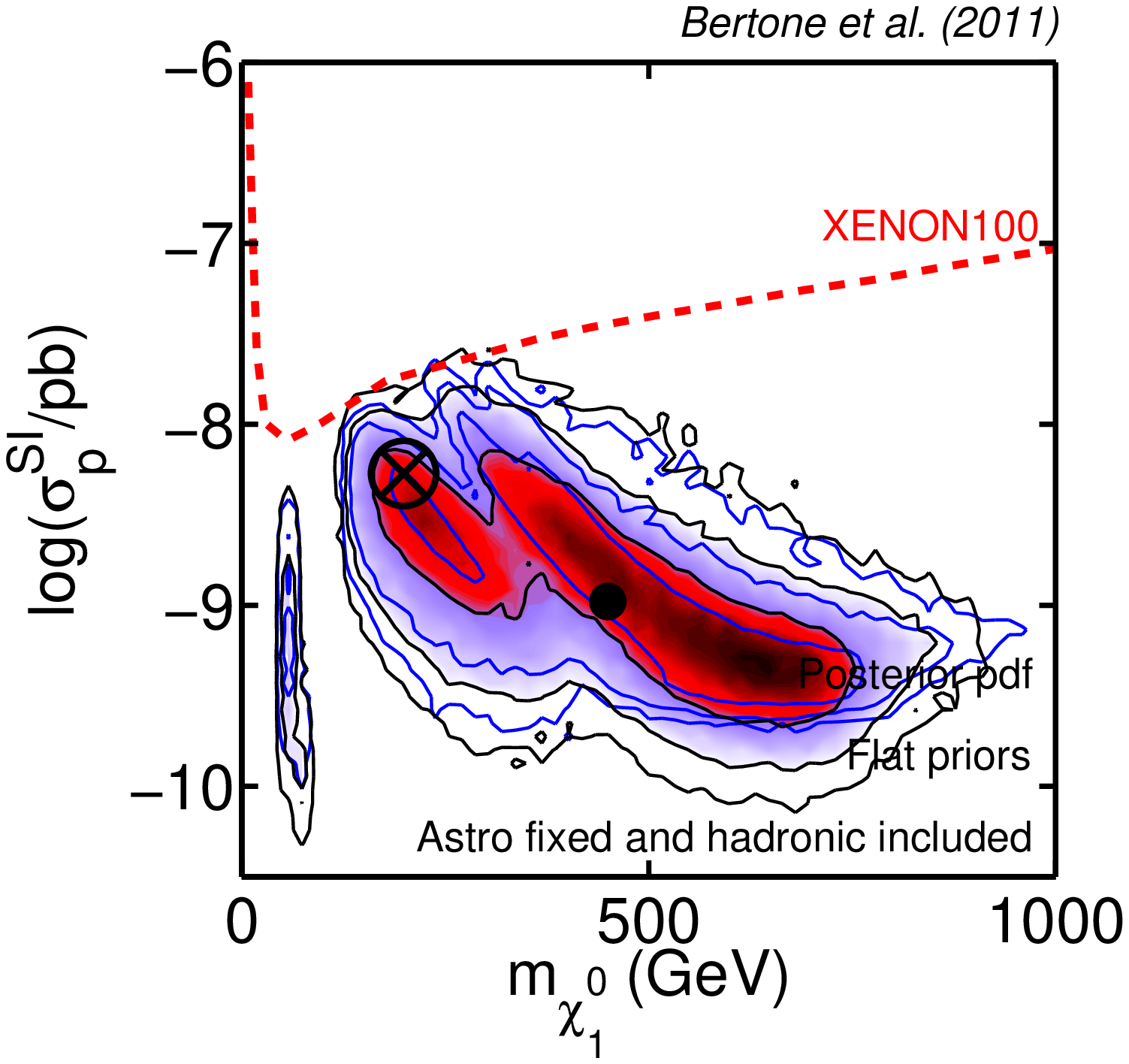} 
\includegraphics[width=0.32\linewidth]{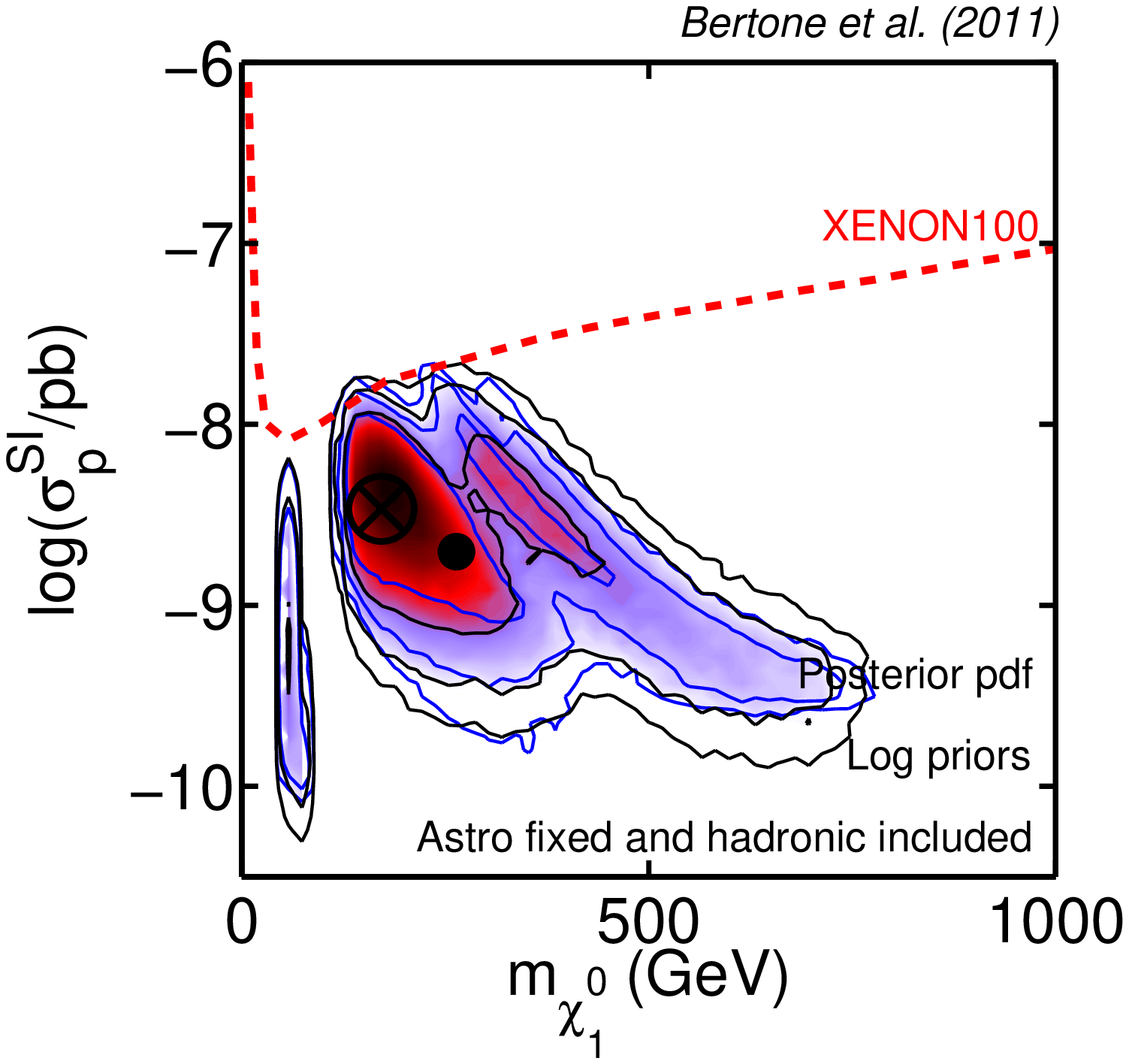}
\includegraphics[width=0.32\linewidth]{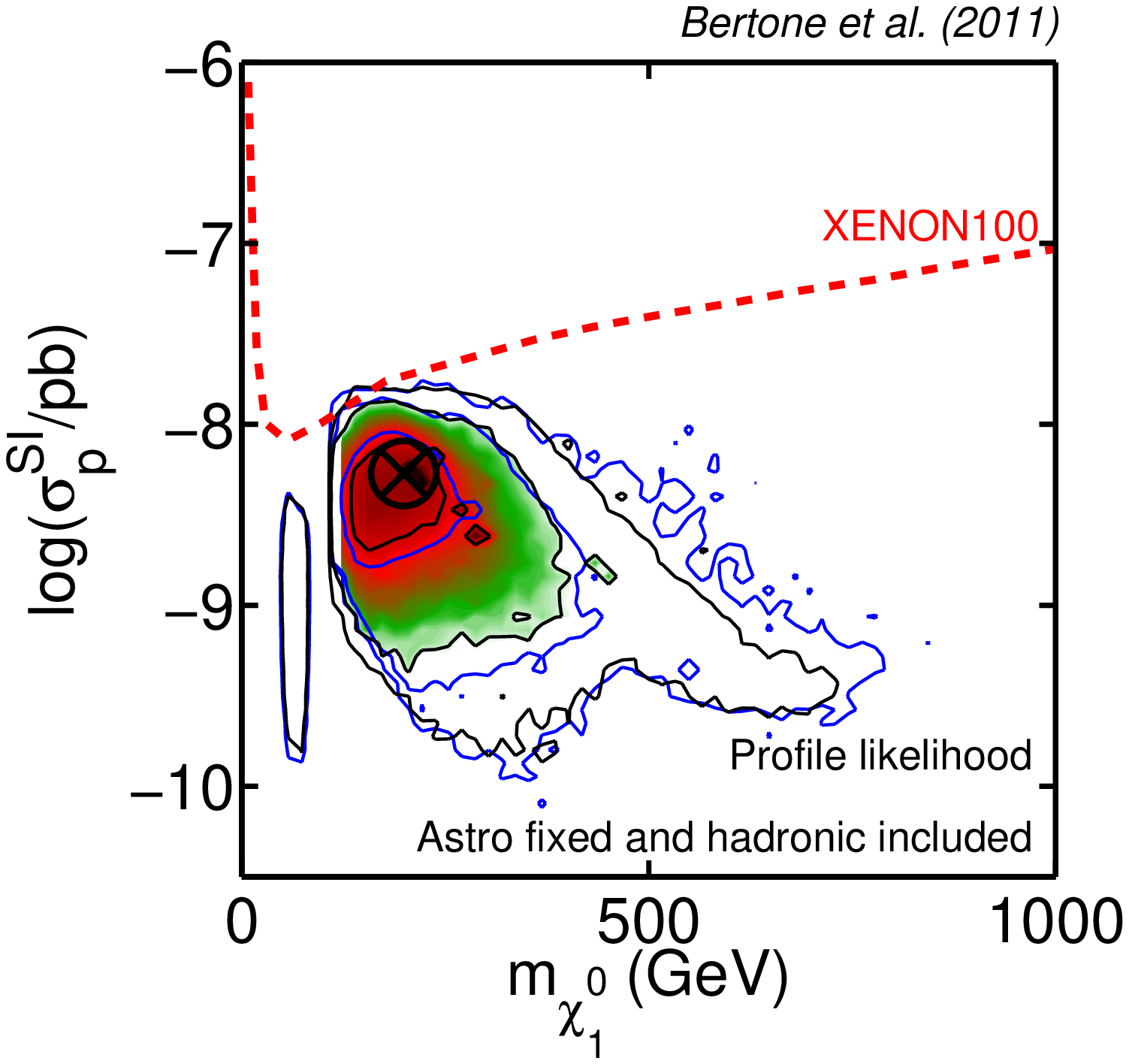}  \\
\includegraphics[width=0.32\linewidth]{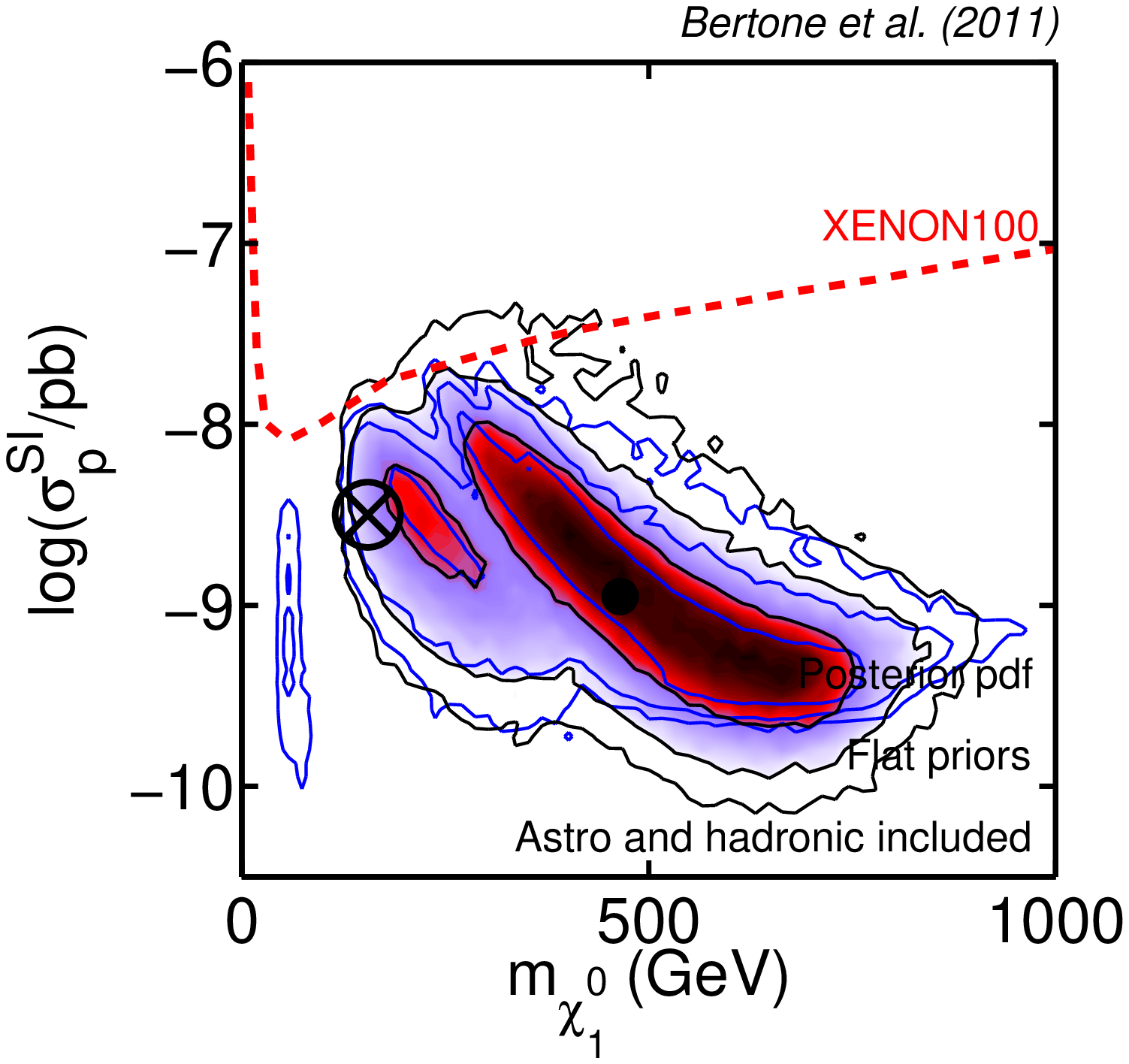} 
\includegraphics[width=0.32\linewidth]{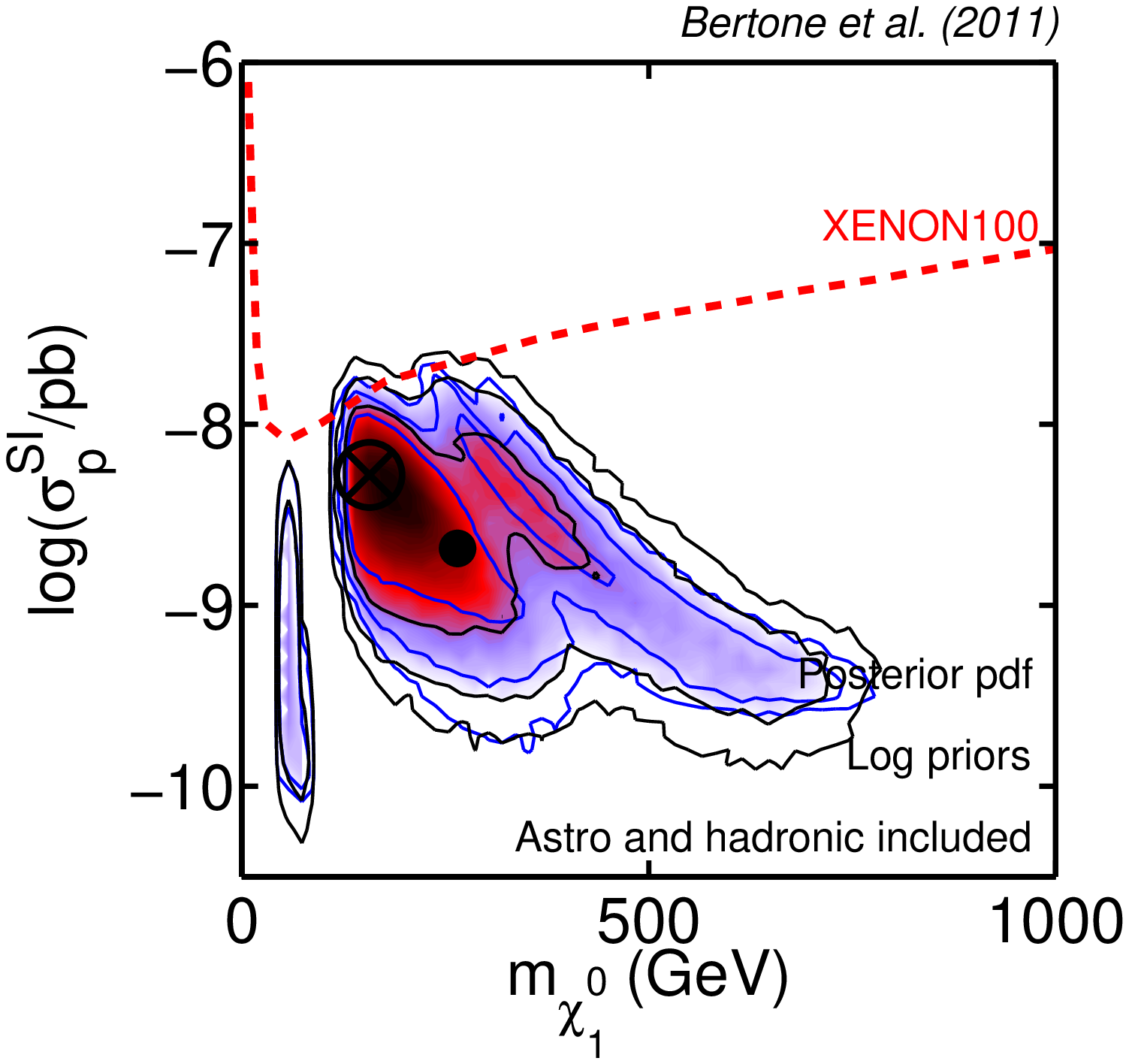}
\includegraphics[width=0.32\linewidth]{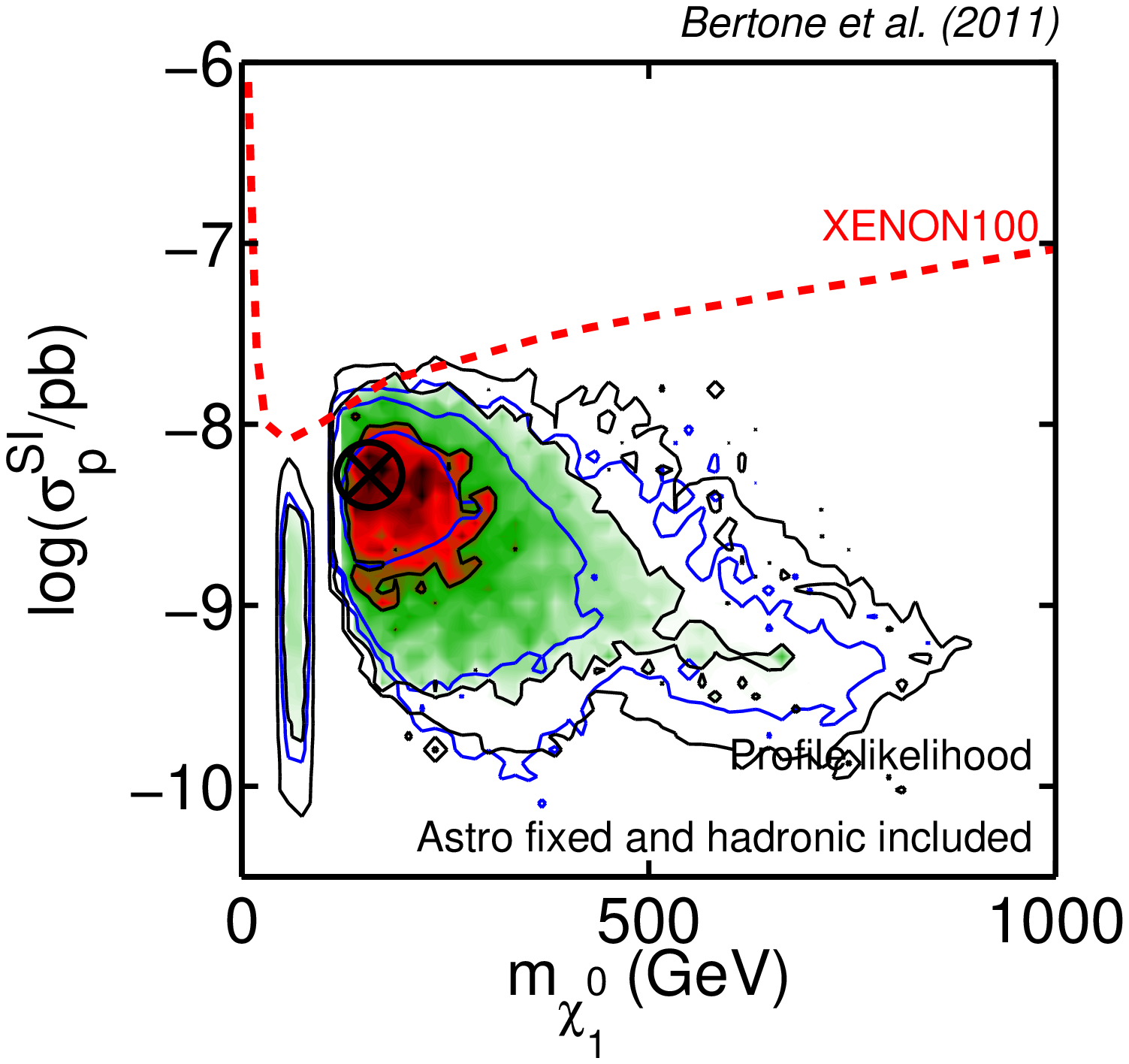} 
\caption{\fontsize{9}{9} \selectfont Top row: Impact of marginalizing (profiling) over hadronic uncertainties as nuisance parameters in the posterior pdf (profile likelihood, right-most panels). Parameters describing astrophysical uncertainties have been fixed to their fiducial values. Bottom row: Impact of further marginalizing/profiling over astrophysical uncertainties. For comparison, the blue contours are the case where both the hadronic and astrophysics nuisance parameters are fixed to their fiducial value. In all panels, all available data have been applied (including LHC and XENON100). The reach of the future ton-scale XENON1T experiment is also indicated (bottom red/dashed line). \label{fig:hadronic_astro_impact}} 
\end{figure*}

\subsection{Influence of hadronic and astrophysical uncertainties.}


We present the effect of including hadronic uncertainties in the top panels of Fig.~\ref{fig:hadronic_astro_impact}. A comparison with the case where the hadronic matrix elements are fixed (grey contours in the figure) shows that although a full analysis leads to slightly more conservative results by increasing slightly the extent of the favoured regions, the simplified analysis with fixed hadronic parameters is accurate enough to draw fairly accurate conclusions on the impact of various searches on the cMSSM parameter space. 

Finally, including both sets of nuisance parameters and marginalizing/profiling over them is the most robust way of applying direct detection constraints. This case is presented in the bottom panels of Fig.~ \ref{fig:hadronic_astro_impact}. Again, we observe a slight broadening of the contours for the Bayesian pdf, compared to the case when both sets of nuisance parameters are kept fixed, but the overall impact on the favoured region is quite small. Regarding the profile likelihood depicted in the bottom-right panel of Fig.~\ref{fig:hadronic_astro_impact}, the profiled contours for the case including nuisance parameters are actually slightly smaller (at 99\%) than when the nuisance parameters are held fixed. This is a consequence of the slightly better value for the best fit $\chi^2$ found in our scan including nuisance parameters (cf Table~\ref{tab:chisquare}), compared with the overall best fit when the nuisance parameters are fixed ($\Delta\chi^2 = 0.05$). A detailed analysis reveals that the improvement is largely ascribable to the better $\gmtwo$ value found in the former scan, which is essentially exactly identical to the experimental central value. For the profile likelihood, even a very small improvement in the best fit value has an impact on the contours as far out in the tails as 99\%, as those are defined with respect to the best fit $\chi^2$. However, given the numerical uncertainties associated with any scan, we can safely conclude that this tightening of the contours is a spurious effect and that the extent of the 99\% region remains qualitatively the same when including both hadronic and astrophysical nuisance parameters in the scan.

\section{Updated prospects for cMSSM discovery} 
\label{sec:prospects}

\begin{figure*}
\centering
\includegraphics[width=0.194\linewidth]{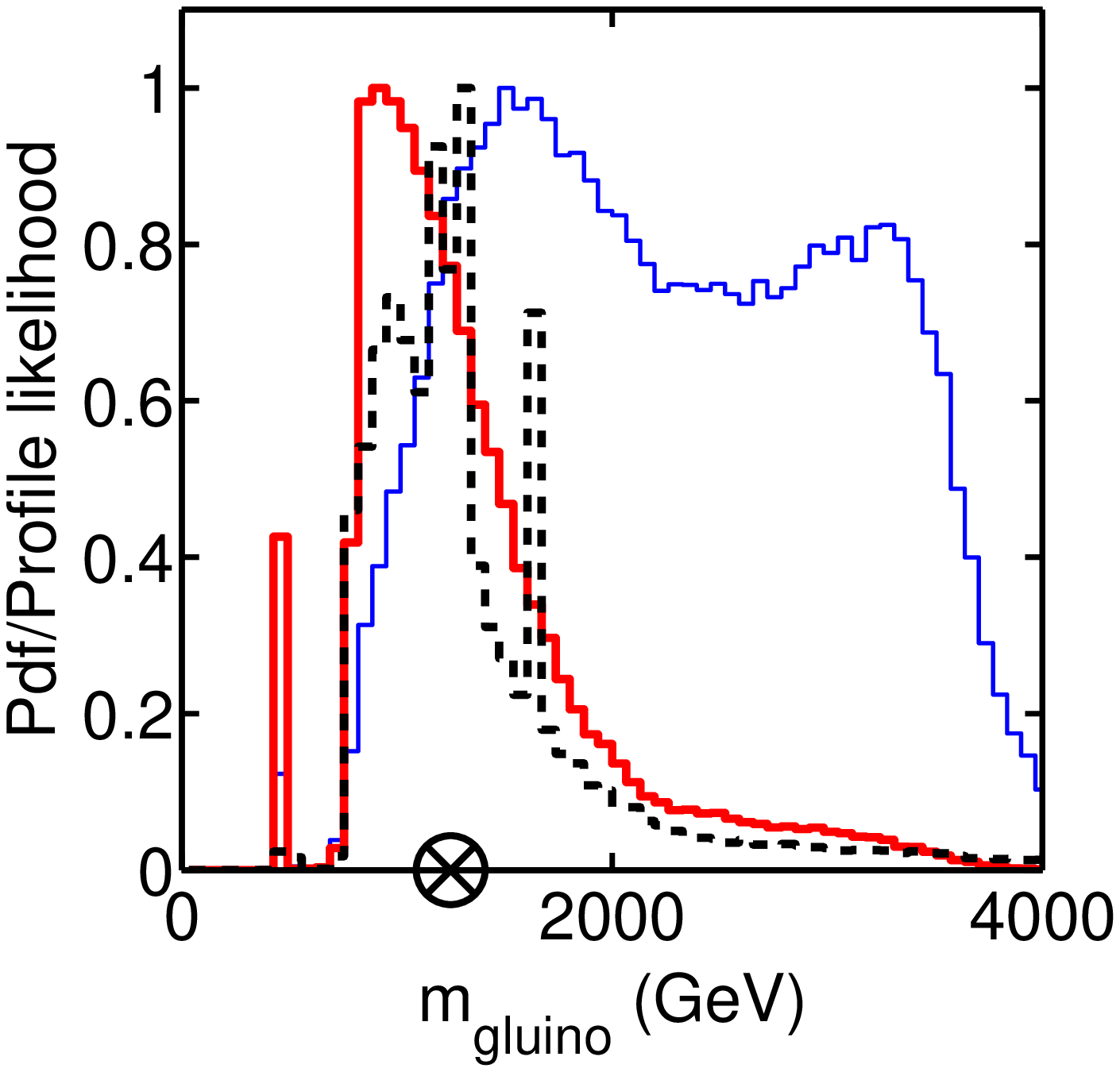}
\includegraphics[width=0.194\linewidth]{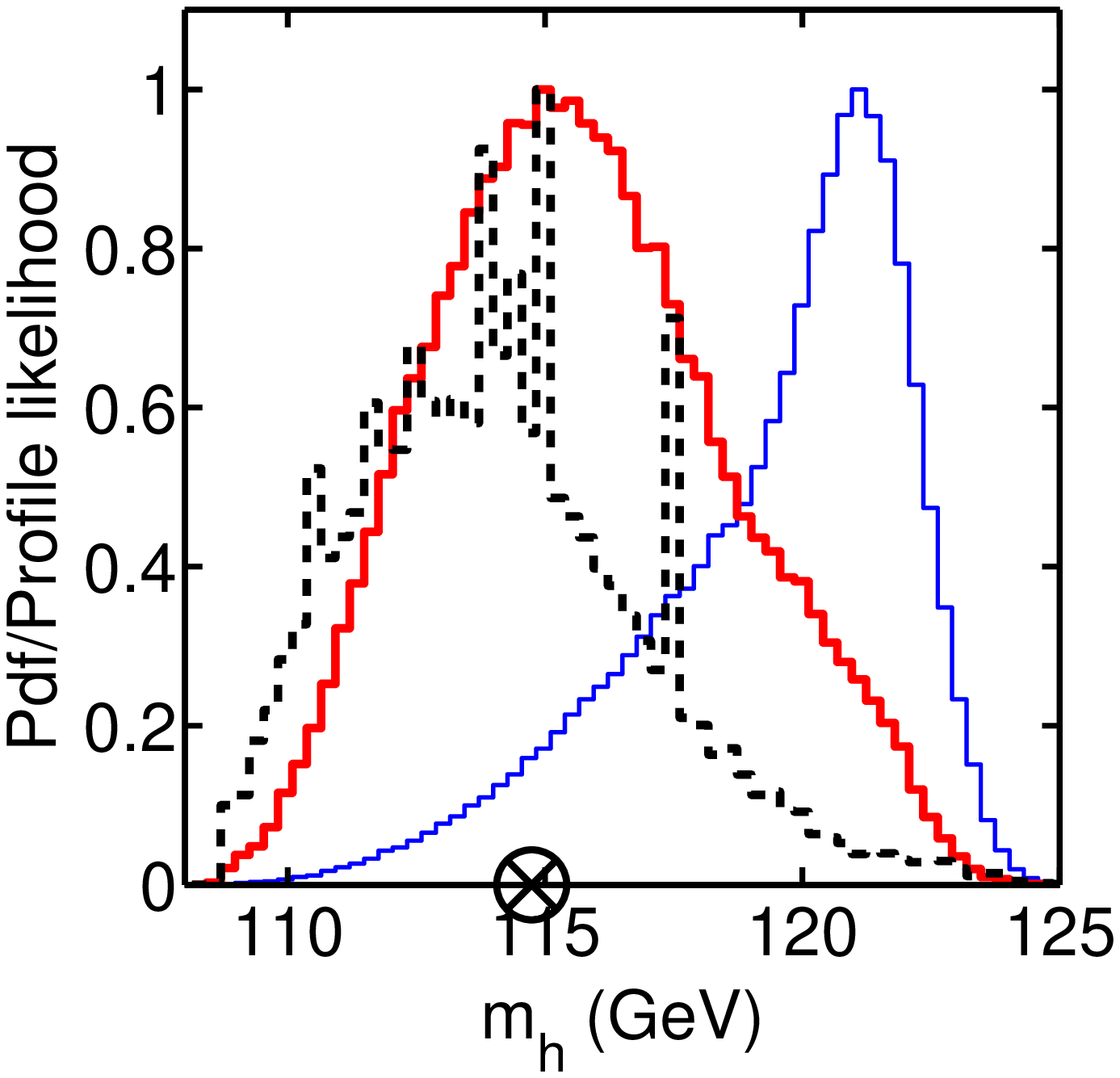}
\includegraphics[width=0.194\linewidth]{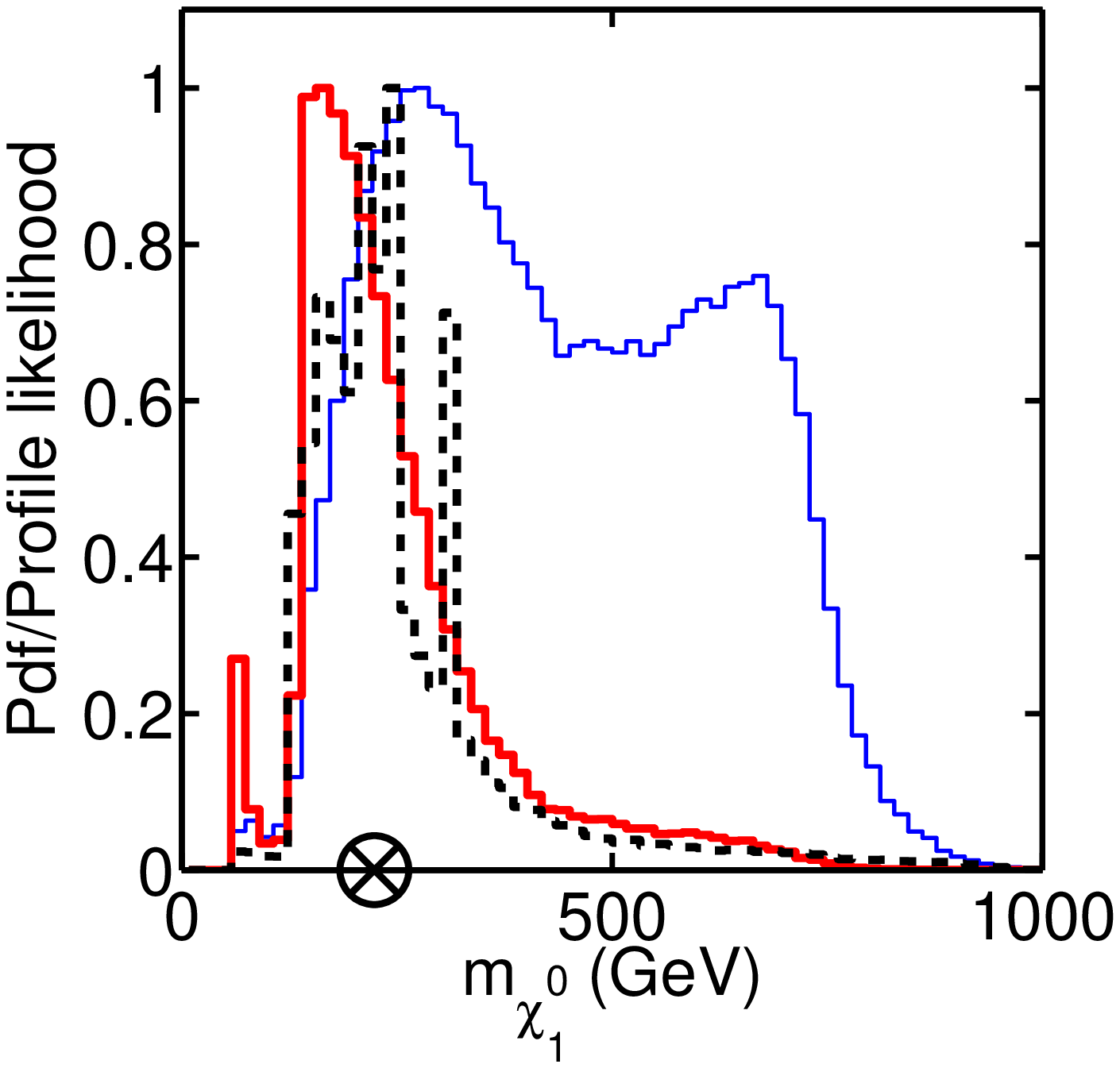}
\includegraphics[width=0.194\linewidth]{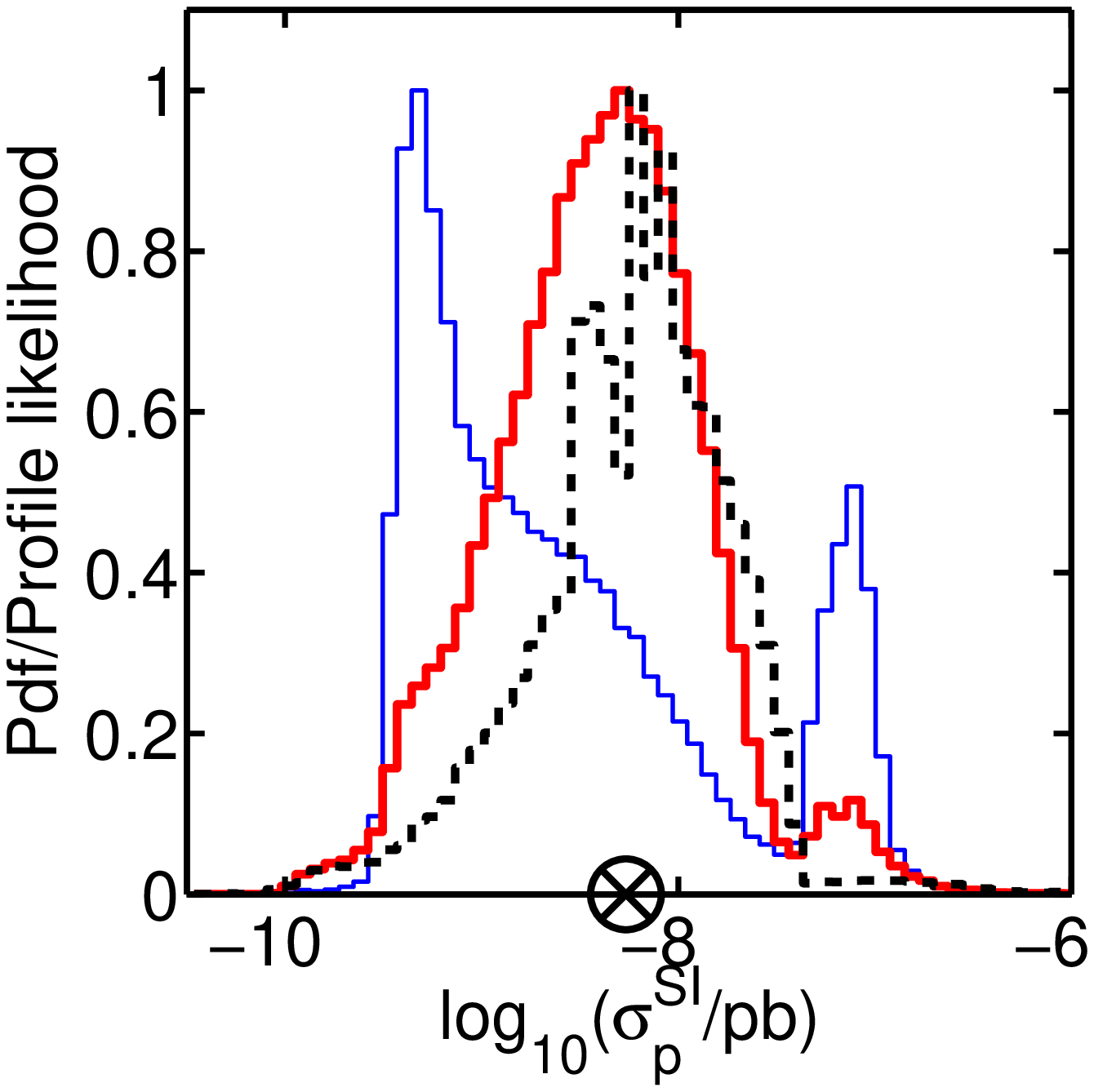}
\includegraphics[width=0.194\linewidth]{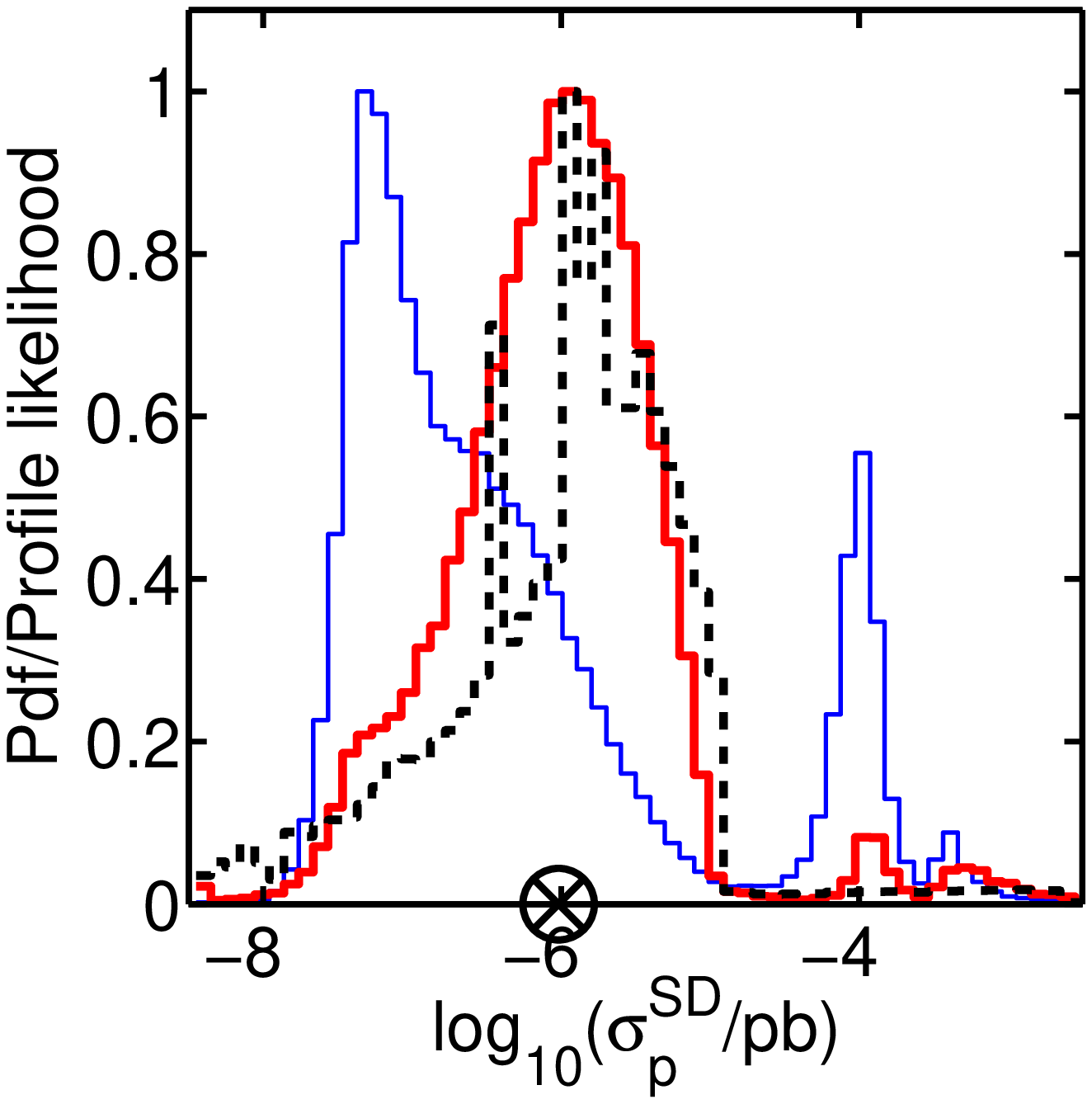} \\ 
\includegraphics[width=0.194\linewidth]{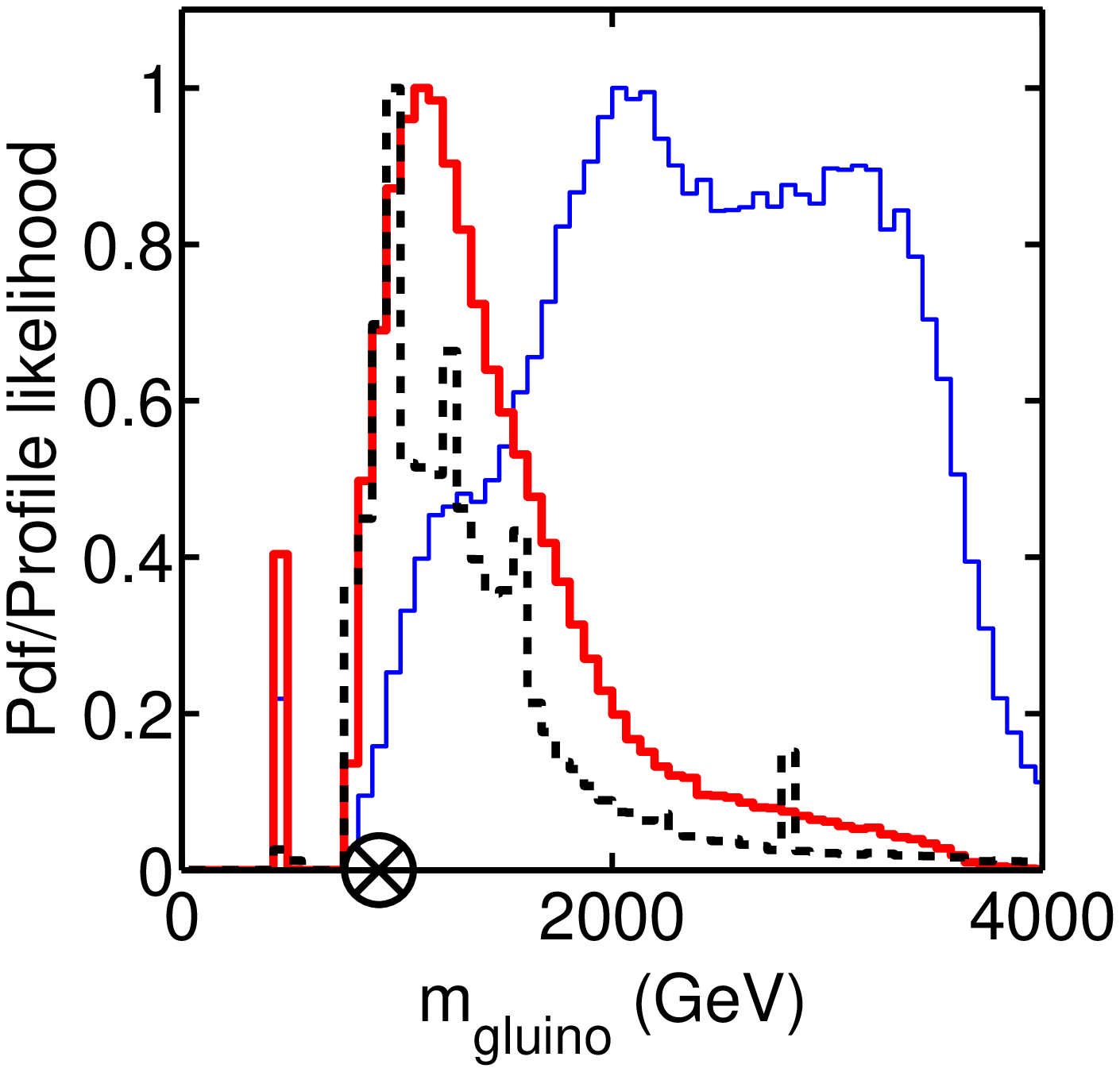}
\includegraphics[width=0.194\linewidth]{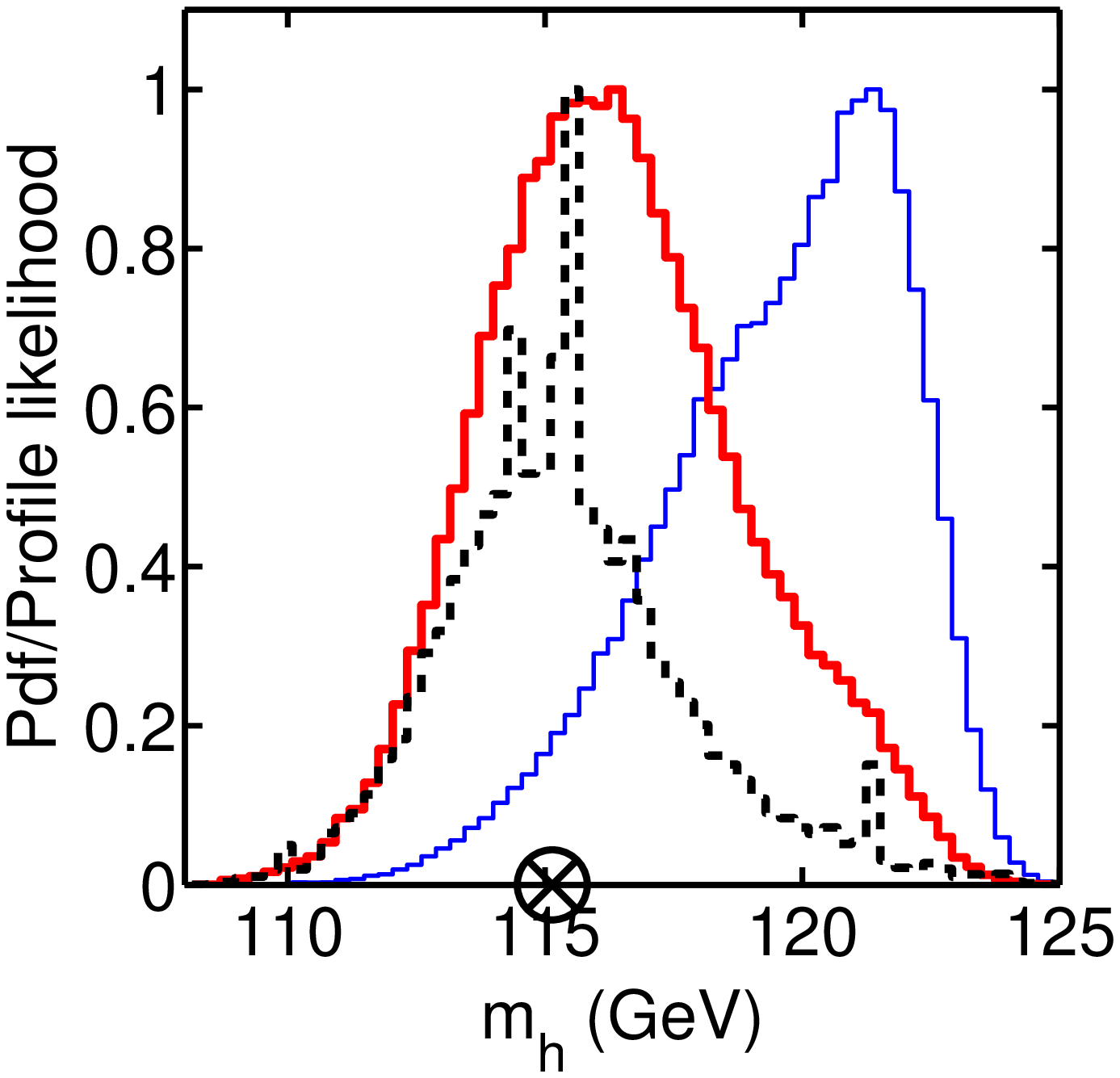}
\includegraphics[width=0.194\linewidth]{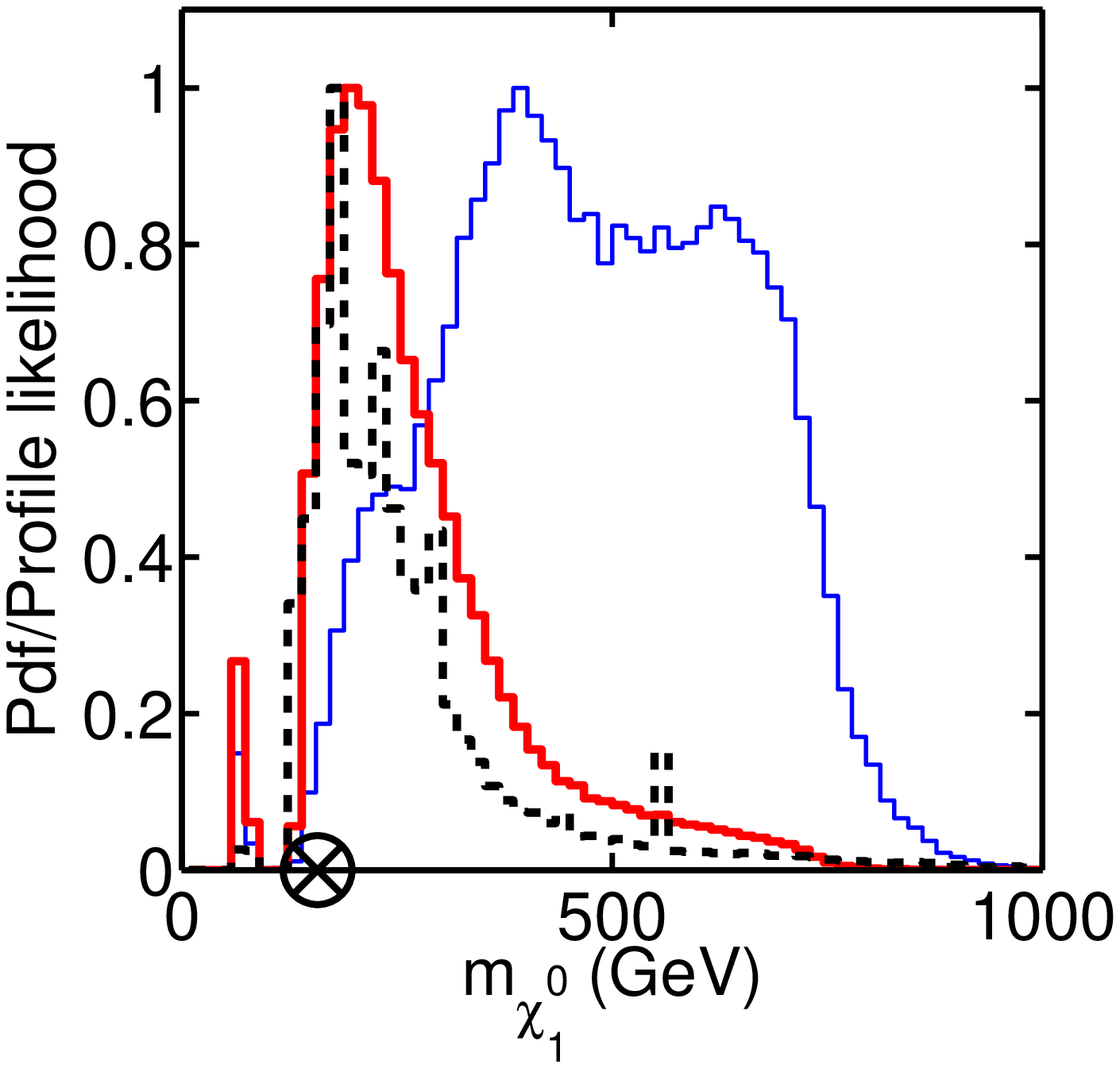}
\includegraphics[width=0.194\linewidth]{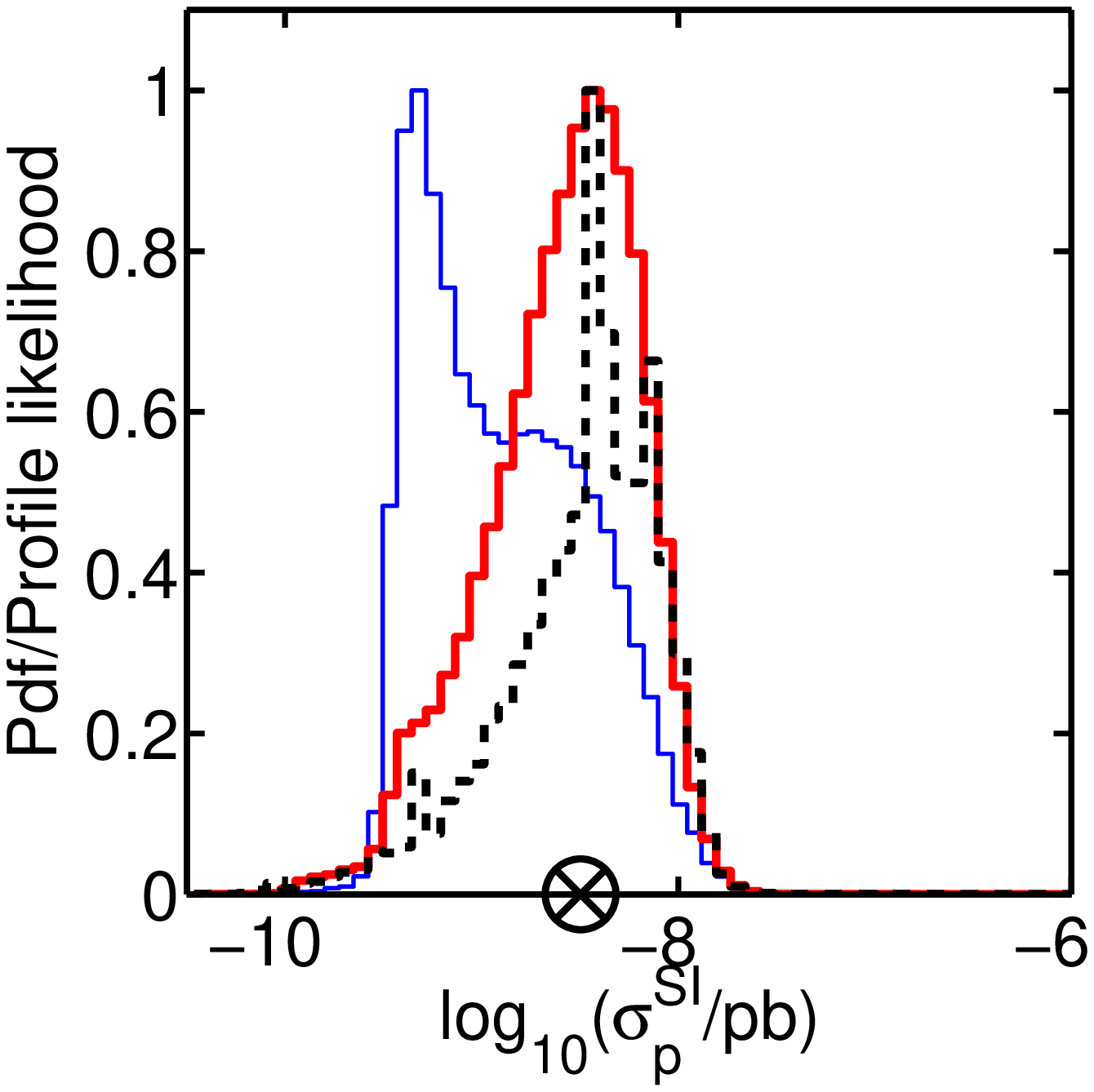}
\includegraphics[width=0.194\linewidth]{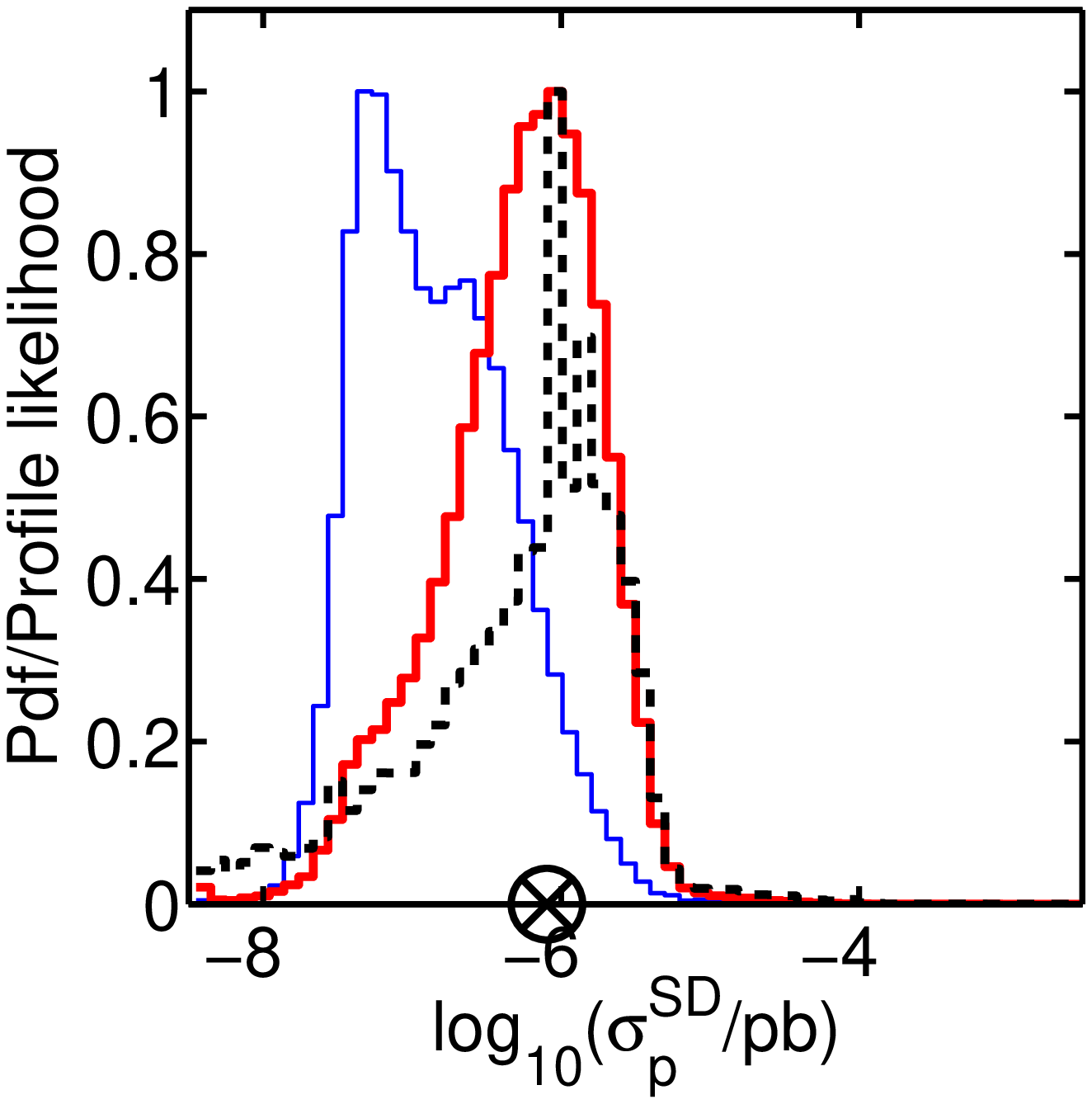} \\
\includegraphics[width=0.194\linewidth]{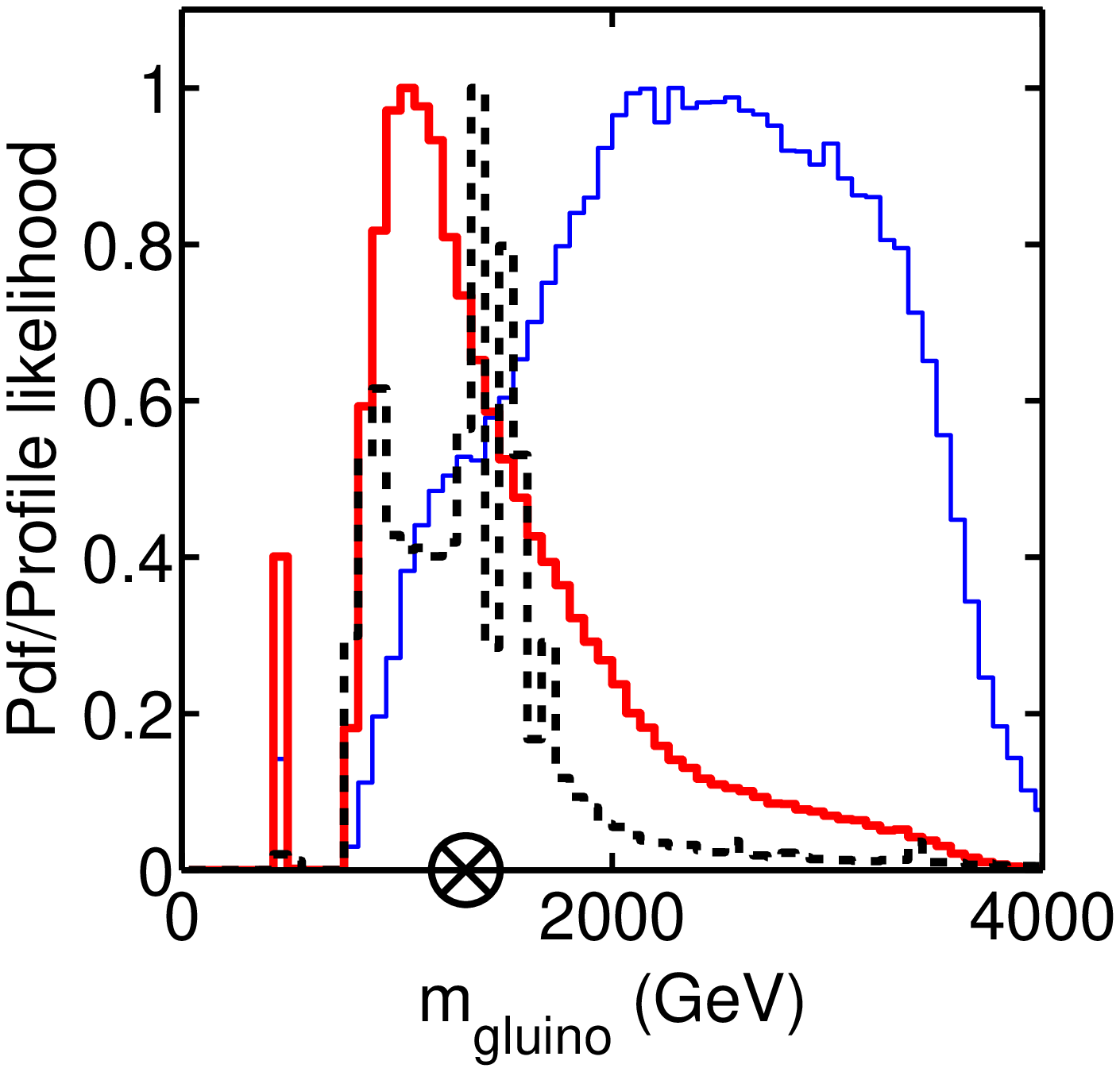}
\includegraphics[width=0.194\linewidth]{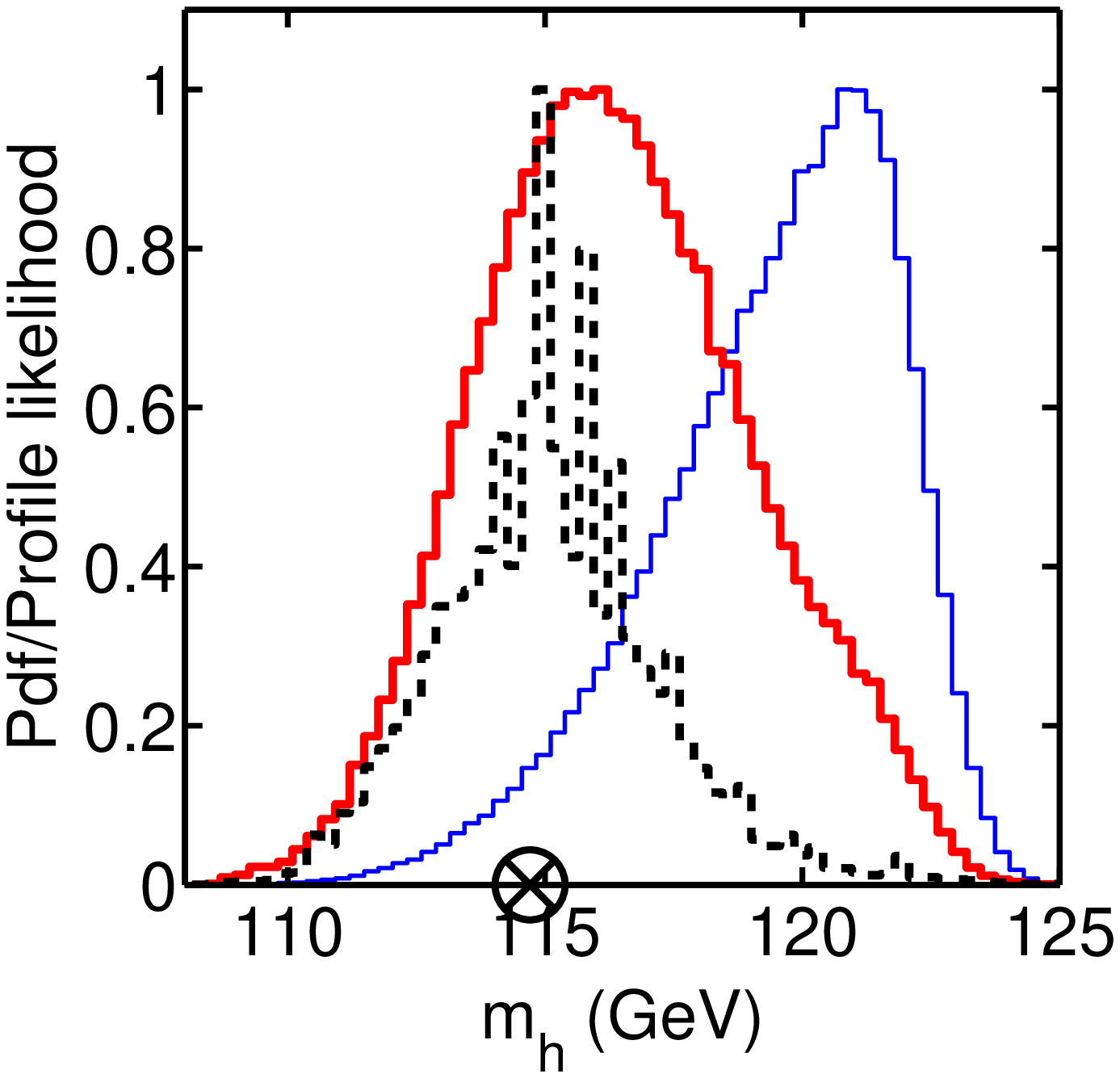}
\includegraphics[width=0.194\linewidth]{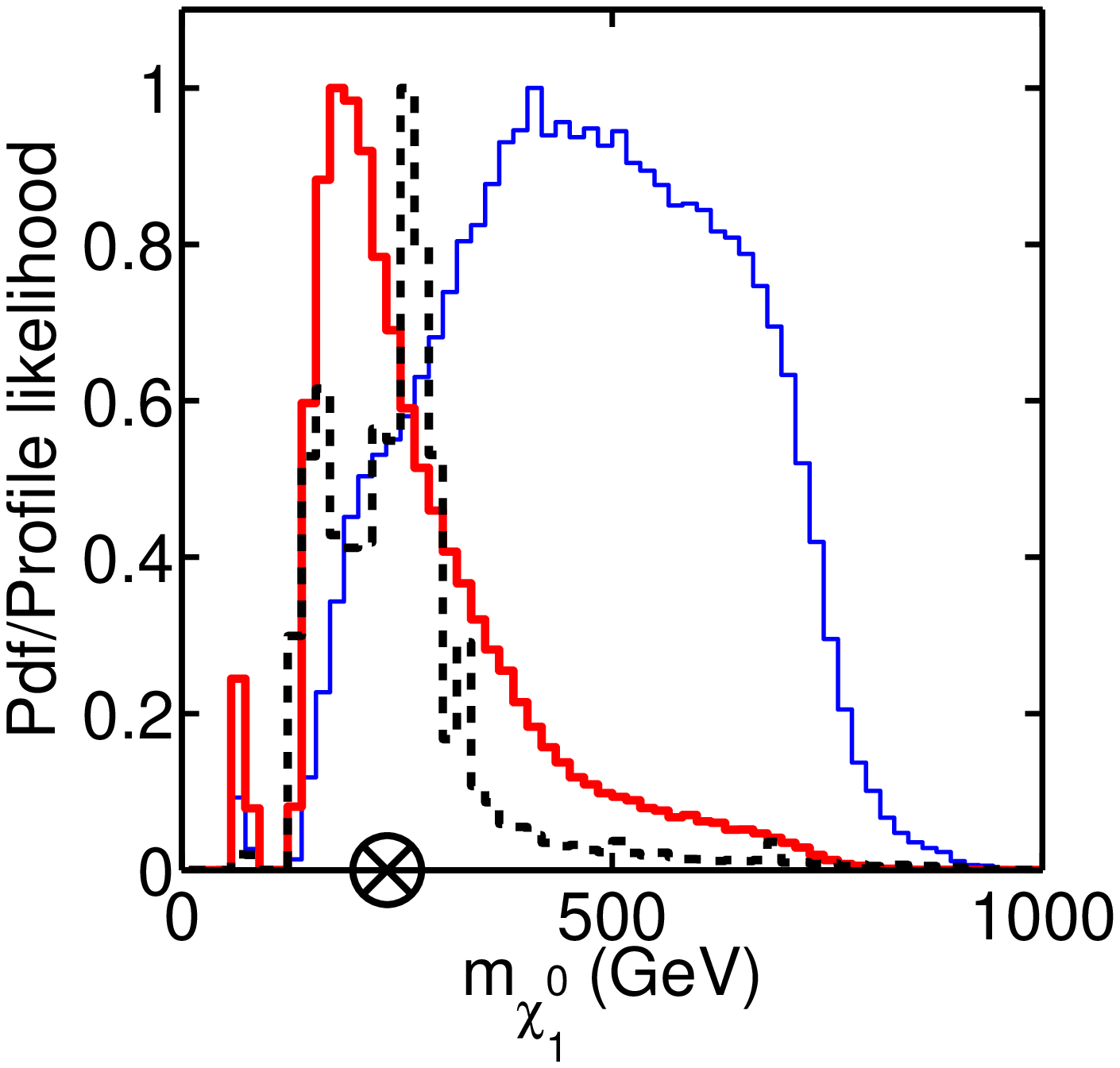}
\includegraphics[width=0.194\linewidth]{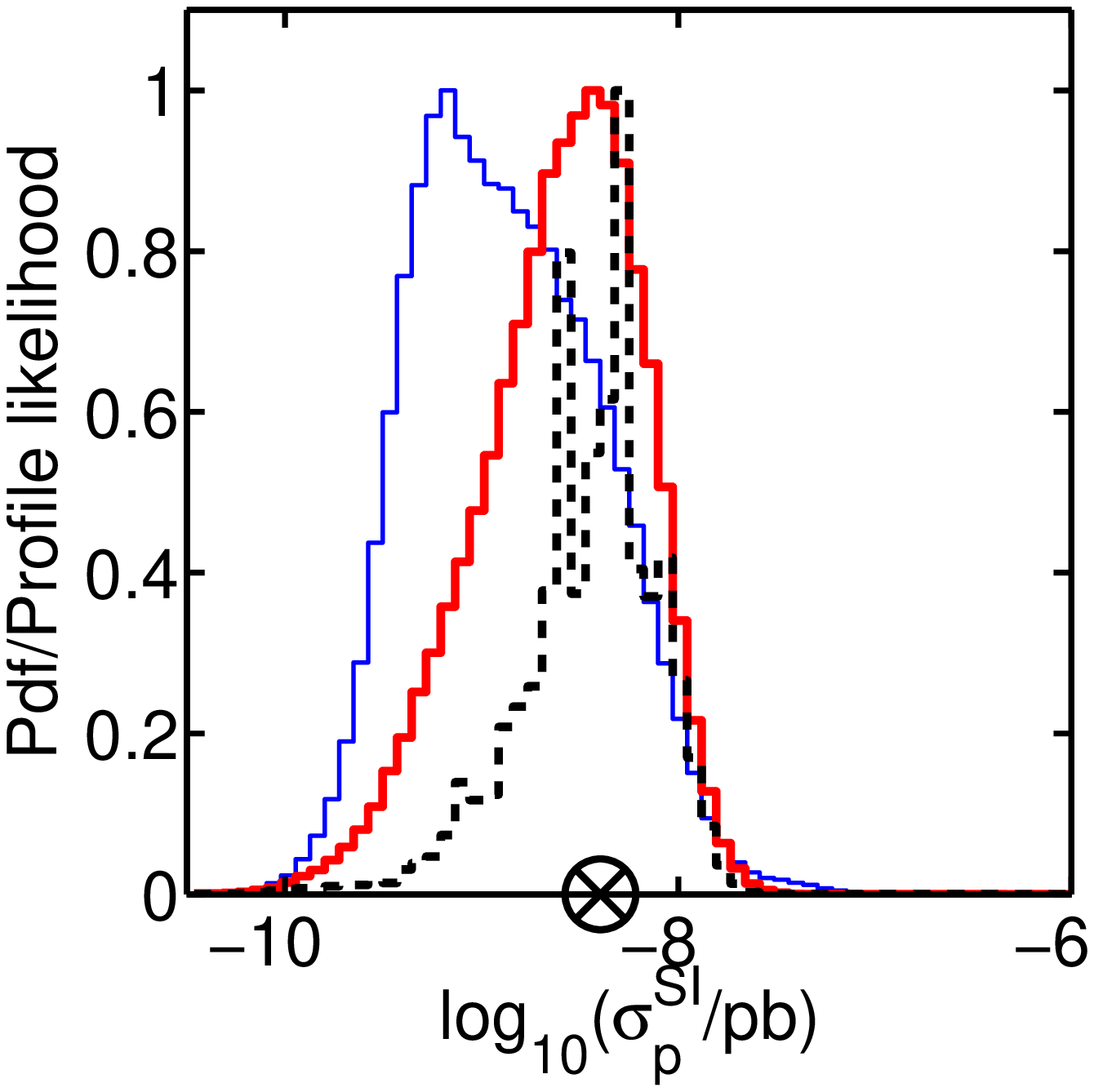}
\includegraphics[width=0.194\linewidth]{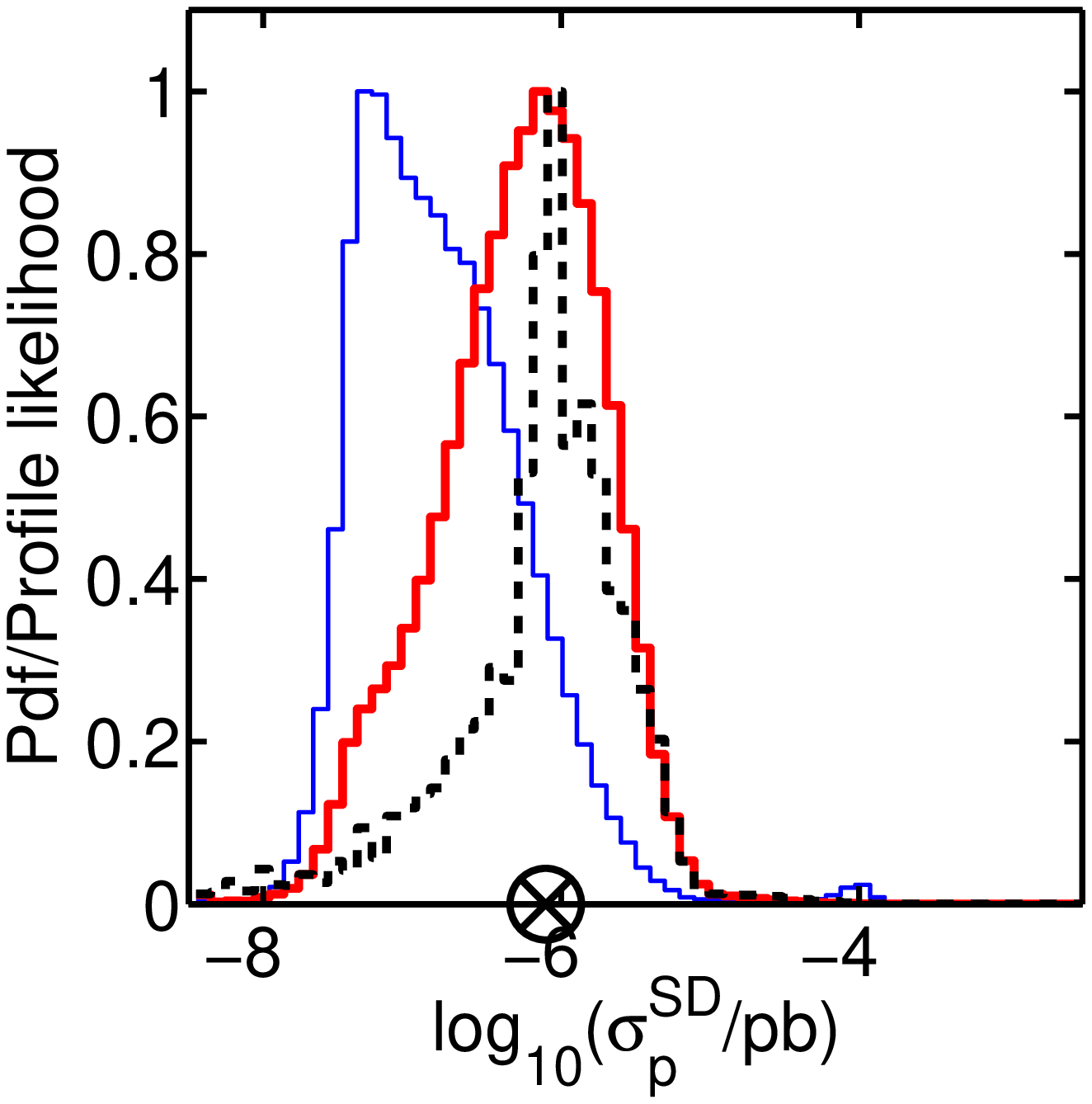}
\caption{\fontsize{9}{9} \selectfont 1D marginal pdf for flat priors (thin solid/blue), log priors (thick solid/red) and 1D profile likelihood (dashed/black) for the gluino mass, the lightest Higgs mass, the neutralino mass, the SI and the SD scattering cross section (from left to right). Top panels include all data except XENON100, middle panels include XENON100 data and bottom panels further marginalize over hadronic and astrophysical uncertainties. The best fit point is indicated by the encircled black cross. \label{fig:prospects}} 
\end{figure*}

We now turn to discuss the implications of our results for detection prospects at the LHC, and via direct and indirect detection channels. Fig.~\ref{fig:prospects} shows 1D posterior distributions and profile likelihoods for the lightest Higgs mass, $m_h$, the gluino mass $m_{\tilde{g}}$, the neutralino mass $m_{\chi^0_1}$, the spin-independent scattering cross section $\sigmaSI$ and the spin-dependent scattering cross section of neutralinos off protons $\sigmaSD$. 

We notice that the 1D marginal distributions for the log prior scan (thick solid/red) and the 1D profile likelihood (dashed/black) are very similar, while the 1D distribution from the flat prior scan (thin solid/blue) still shows some residual volume effect. This manifests itself e.g.~in the shift of the bulk of the probability density to larger gluino and neutralino masses. However, a robust result of our scans is that the best fit neutralino mass is fairly small (in the range $\sim 150 - 250$ GeV), the lightest Higgs very light (just above the LEP exclusion limit), and the spin-independent scattering cross section a mere factor of $\sim 2$ below current XENON100 limits. This therefore puts our best fit point for the cMSSM easily within reach of the next generation of direct detection experiments. In particular, the upcoming scaled-up version of XENON100, XENON1T, is expected to probe by 2015 practically the entire 99\% posterior and profile likelihood shown above, reaching a sensitivity better than $\sigmaSI =10^{-10}$~pb in a mass range extending from 20 to 300 GeV (see e.g. the recent assessment of the XENON1T reach in Ref. \cite{xenon1T}). 

Interestingly, the FP is a region that presents promising prospects for {\it indirect} DM searches, especially for what concerns the detection of high-energy neutrinos from DM annihilations in the Sun (e.g. Ref. \cite{Baer:2005ky,Trotta:2009gr}), that should be possible with the IceCube neutrino telescope if the $\sigmaSD$ is larger than $\approx 3 \times 10^{-5}$~pb (for an observation time of 5 years, and including the densely instrumented region DeepCore at the center of IceCube, see e.g. Ref. \cite{Halzen:2009vu} for details). The fact that XENON100 rules out the FP branch has therefore a negative impact on the prospects for indirect detection, and further constrains the possibility to probe DM in the cMSSM with astrophysical experiments. Upon inspection of the posterior pdf and profile likelihood in the right-most panels of Fig.~\ref{fig:prospects}, we see that after inclusion of XENON100 data the region with the highest pdf and profile likelihood values falls below the sensitivity of IceCube, although a detection is still statistically possible in the tails of the distributions but now more unlikely. Finally, comparison of the middle and bottom rows of Fig.~\ref{fig:prospects} shows that inclusion of the astrophysical and hadronic uncertainties does not change the overall conclusions dramatically, with only mild shifts to the resulting inferences resulting from inclusion of the extra nuisance parameters in the scan (bottom row).

\section{Conclusion}
\label{secconclusion}

We have presented in this paper new global fits of the cMSSM, including the most recent constraints from the LHC and the XENON100 experiment. Besides the uncertainties on Standard Model quantities, our analysis takes into account both astrophysical and hadronic uncertainties that enter in the calculation of the rate of WIMP-induced recoils in direct detection experiment. We have shown that the FP branch of the cMSSM is robustly ruled out, even when all uncertainties are taken into account. These results highlight the complementarity of direct detection experiments to collider searches. Indeed, direct detection data would still be required to rule out the FP even if the LHC exclusion limits are extended to a total integrated luminosity of 7 fb$^{-1}$ at 7 TeV center-of-mass energy~\cite{Bechtle:2011dm}, which might be achieved by the end of 2012.

Although our results are specific to the cMSSM, we note that the conditions for the occurrence of the FP can actually be extended to a
general class of unconstrained Supersymmetric scenarios. In models where
no universality assumptions on the soft mass parameters are made, the FP region
can be shifted to either larger or lower scalar mass parameters, but
the general feature of the smallness of the $\mu$ parameter is
preserved, as are its implications for dark matter. In Ref.~\cite{Baer:2005ky}, for example, a low-energy parametrization of the FP was
studied that could be used to describe this region irrespectively of
the high-energy assumptions. Although the analysis of these general constructions is beyond the scope of the
present work, it seems reasonable to extend our conclusions to the general case, given 
the similarity of the predicted direct detection rates.

While this work was under completion, Refs.~\cite{Farina:2011bh,Buchmueller:2011ki} performed a similar analysis of the impact of LHC and XENON100 data on the parameter space of several theoretical models, including the cMSSM. While our results about the viability of the FP regions are qualitatively similar, our analysis differs from that in the above references in that we studied the impact of both hadronic and astrophysical uncertainties (both neglected in Ref.~\cite{Farina:2011bh}, while Ref.~\cite{Buchmueller:2011ki} includes only hadronic uncertainties). Furthermore, we adopt a more sophisticated statistical framework than Ref.~\cite{Farina:2011bh} (where the parameter space was explored using random scans, rather than the more sophisticated statistical scans employed here) and perform a detaild, quantitative comparison between the Bayesian and profile likelihood results (absent in Ref.~\cite{Buchmueller:2011ki}). 


\acknowledgments  
R.T.~would like to thank AIMS, the University of Zurich and the Kavli Institute for Theoretical Physics at the UCSB for hospitality. C.S. is partially supported by a scholarship of the ``Studienstiftung des deutschen Volkes''.
D.G.C. is supported by the Ram\'on y Cajal program of the Spanish MICINN and also thanks the support of the MICINN under grant FPA2009-08958, the Community of Madrid under grant HEPHACOS S2009/ESP-1473, and the European Union under the Marie Curie-ITN program PITN-GA-2009-237920. We thank the support of the Consolider-Ingenio 2010 Programme under grant MultiDark CSD2009-00064. This research was supported in part by the National Science Foundation under Grant No. NSF PHY05-51164.
We also thank Henrique Araujo, Chiara Arina, Laura Baudis, Eilam Gross, Joakim Edjs\"o, Christopher Savage and Pietro Slavich for useful discussions.

\end{document}